\documentclass[pra, twocolumn, letterpaper,superscriptaddress]{revtex4}

\usepackage{amsmath}
\usepackage{amssymb}
\usepackage{mathrsfs}
\usepackage[mathscr]{euscript}	%For nice fonts with \mathscr
\usepackage{xspace}
\usepackage{graphicx}				%Figures
\usepackage{braket}					%Dirac braket notation
\usepackage[utf8]{inputenc}
\usepackage{enumitem}			%for itemized with no margins
\usepackage{color} 

\usepackage{ifpdf}
\ifpdf
\fi

\setcitestyle{numbers,square}
\usepackage{hyperref}
\hypersetup{
  colorlinks   	= true, 	%Colours links instead of ugly boxes
  urlcolor     	= blue, 	%Colour for external hyperlinks
  linkcolor    	= blue, 	%Colour of internal links
  citecolor   	= blue 	%Colour of citations
}
\newcommand{\eq}[1]{(\ref{#1})}
\newcommand{\Eq}[1]{Eq.~(\ref{#1})}
\newcommand{\Eqs}[1]{Eqs.~(\ref{#1})}
\newcommand{\Fig}[1]{Fig.~\ref{#1}}
\newcommand{\Sec}[1]{Sec.~\ref{#1}}
\newcommand{\Ref}[1]{Ref.~\cite{#1}}
\newcommand{\Refs}[1]{Refs.~\cite{#1}}
\newcommand{\App}[1]{Appendix~\ref{#1}}

\newcommand{\eg}{{e.g.,\/}\xspace}
\newcommand{\ie}{{i.e.,\/}\xspace}

\newcommand{\pd}{\partial}
\newcommand{\del }{\vec{\nabla}}

\newcommand{\hc}{\text{h.\,c.}}

\newcommand{\moysin}[1]{	\{ \! \{	 #1 \} \!	\} 	}

\newcommand{\mc}[1]{\mathcal{#1}}
\newcommand{\mcc}[1]{\mathfrak{#1}}

\newcommand{\mcu}[1]{\mathscr{#1}}
\newcommand{\oper}[1]{\hat{\mcu{#1}} }
\renewcommand{\vec}[1]{{\boldsymbol{\rm #1}}}

\newcommand{\avg}[1]{ \boldsymbol{\langle} \!\langle  #1  \rangle\! \boldsymbol{\rangle} }

\begin{document}

\title{Ponderomotive dynamics of waves in quasiperiodically modulated media}

\begin{abstract}
Similarly to how charged particles experience time-averaged ponderomotive forces in high-frequency fields, linear waves also experience time-averaged refraction in modulated media. Here we propose a covariant variational theory of this ``ponderomotive effect on waves'' for a general nondissipative linear medium. Using the Weyl calculus, our formulation accommodates waves with temporal and spatial period comparable to that of the modulation (provided that parametric resonances are avoided). Our theory also shows that any wave is, in fact, a polarizable object that contributes to the linear dielectric tensor of the ambient medium. The dynamics of quantum particles is subsumed as a special case. As an illustration, ponderomotive Hamiltonians of quantum particles and photons are calculated within a number of models. We also explain a fundamental connection between these results and the commonly known expression for the electrostatic dielectric tensor of quantum plasmas.
\end{abstract}

\author{D.~E. Ruiz}
\affiliation{Department of Astrophysical Sciences, Princeton University, Princeton, New Jersey 08544, USA}
\author{I.~Y. Dodin}
\affiliation{Department of Astrophysical Sciences, Princeton University, Princeton, New Jersey 08544, USA}
\affiliation{Princeton Plasma Physics Laboratory, Princeton, New Jersey 08543, USA}

\date{\today}

\maketitle

%%%%%%%%%%%%%%%%%%%%%%%%%%%%%%%%%%%%%%%%%%%%%%
%%%%%%%%%%%%%%%%%%%%%%%%%%%%%%%%%%%%%%%%%%%%%%
\section{Introduction}

It is well known that a non-uniform high-frequency electromagnetic (EM) field can produce a time-averaged force, known as the ponderomotive force, on any particle that is charged or, more generally, has nonzero polarizability \cite{Boot:1957eo,Gaponov:1958un,ref:dewar72c,Cary:1977ud,Grebogi:1979gd,Dodin:2008fe,Brodin:2010bc}. This effect has permitted a number of applications ranging from atomic cooling to particle acceleration \cite{Ashkin:1970ga,Malka:1996zz}, but many other interesting opportunities remain. In particular, similar manipulations can be practiced on waves too. As shown recently in \Ref{Dodin:2014iz}, any wave propagating through a temporally and (or) spatially modulated medium generally experiences time-averaged refraction determined by the modulation intensity \cite{foot:intro}. It was also shown in \Ref{Dodin:2014iz} that this ``ponderomotive effect on waves'' subsumes the ponderomotive dynamics of particles as a special case because, quantummechanically, particles can be represented as waves. However, \Ref{Dodin:2014iz} assumes that the wave period (both temporal and spatial) is much smaller than the modulation period. This approximation limits the applicability of the theory. One may wonder then whether it can be relaxed (without specifying the type of waves being considered) and whether new interesting physics can be discovered then.

Here we answer these questions positively by proposing a general theory of the ponderomotive effect on waves. In contrast with \Ref{Dodin:2014iz}, this theory can describe waves with temporal and spatial period comparable to that of the modulation (provided that parametric resonances are avoided). Using the Weyl calculus, we explicitly derive the effective dispersion symbol \eq{eq:Deff_final} that governs the time-averaged dynamics of a wave in a quasiperiodically modulated medium. This result is later used to obtain the wave \textit{ponderomotive Hamiltonian} \eq{eq:Heff}. This formulation can be understood as a generalization of the oscillation-center (OC) theory, which is known from classical plasma physics \cite{Dewar:1973jq,Cary:1981tw,Brizard:2009cp}, to any linear waves and quantum particles in particular. Our theory also shows that \textit{any wave is, in fact, a polarizable object} that contributes to the linear dielectric tensor of the ambient medium. As an illustration, ponderomotive energies of quantum particles and photons are calculated within a number of models and compared with simulations. In particular, we find that quantum effects can change the sign of the ponderomotive force. We also explain a fundamental connection between these results and the commonly known expression for the quantum-plasma electrostatic dielectric function. This work also serves as a stepping stone to improving the understanding of modulational instabilities in general wave ensembles, as will be reported separately.

It is to be noted that effective Hamiltonians for temporally driven systems have been studied in condensed matter physics \cite{Grozdanov:1988jy,Gilary:2003bj,Mikami:2016in,Goldman:2015kc,Eckardt:2015hp,Bukov:2015gu,Goldman:2014bva,Novicenko:2016vb}. However, these studies are mainly focused on systems described by the Schr\"odinger equation and use the modulation period as the small parameter. In contrast, we study more general waves and expand in the modulation amplitude rather than period. This way, we can calculate the ponderomotive effect on waves using the Weyl calculus, which provides a direct connection with classical physics and the aforementioned OC theory in particular.

This work is organized as follows. In \Sec{sec:notation} the basic notation is defined. In \Sec{sec:basic} we present the variational formalism and the main assumptions used throughout the work. In \Sec{sec:gen_theory} we derive a general expression for the effective wave action. In \Sec{sec:pondero} we present a theory of ponderomotive dynamics for eikonal waves. In \Sec{sec:applications} we apply the theory to specific examples. In \Sec{sec:mod} we show the fundamental connection between the ponderomotive energy that we derive in this paper and the commonly known dielectric tensor of quantum plasma. In \Sec{sec:conclusions} we summarize our main results. Some auxiliary calculations are presented in the Appendices. This includes an introduction to the Weyl calculus that we extensively use in the paper (\App{app:Weyl}) and details of some of the calculations presented (\App{app:auxiliary}).

%%%%%%%%%%%%%%%%%%%%%%%%%%%%%%%%%%%%%%%%%%%%%%
%%%%%%%%%%%%%%%%%%%%%%%%%%%%%%%%%%%%%%%%%%%%%%
\section{Notation}
\label{sec:notation}

The following notation is used throughout the paper. The symbol ``$\doteq$'' denotes definitions. Unless otherwise specified, natural units are used in this work so that the speed of light equals one ($c = 1$), and so does the Planck constant ($\hbar = 1$). The Minkowski metric is adopted with signature $(+, -, -, -)$. Greek indices span from $0$ to $3$ and refer to spacetime coordinates, $x^\mu=(x^0, \vec{x})$, with $x^0=t$. Also, $\pd_\mu \doteq \pd/ \pd x^\mu=( \pd_t, \del)$, and $\mathrm{d}^4x \doteq \mathrm{d}x^0\,\mathrm{d}^3 \vec{x}$. Latin indices span from $1$ to $3$ and denote the spatial variables, \ie $\vec{x} = (x^1, x^2, x^3)$, and $\pd_i \doteq \pd/\pd x^i$. Summation over repeated indexes is assumed. For arbitrary four-vectors $a$ and $b$, we have: $a \cdot b \doteq a^\mu b_\mu = a^0 b^0 - \vec{a}\cdot \vec{b}$. The Dirac bra-ket notation is used to denote $\ket{\Psi}$ as a state of the Hilbert space defined over $\mathbb{R}^4$. In Euler-Lagrange equations (ELEs), the notation ``$\delta a: $'' denotes that the corresponding equation was obtained by extremizing the action integral with respect to $a$.

%%%%%%%%%%%%%%%%%%%%%%%%%%%%%%%%%%%%%%%%%%%%%%
%%%%%%%%%%%%%%%%%%%%%%%%%%%%%%%%%%%%%%%%%%%%%%
\section{Physical Model}
\label{sec:basic}

%%%%%%%%%%%%%%%%%%%%%%%%%%%%%%%%%%%%%%%%%%%%%%
\subsection{Wave action principle}
\label{sec:assumptions}

We represent a wave field, either quantum or classical, as a scalar complex function $\Psi(x)$. The dynamics of any nondissipative linear wave can be described by the least action principle, $\delta \Lambda =0$, where the real action $\Lambda$ is bilinear in the wave field \cite{Dodin:2014hw}. In the absence of parametric resonances \cite{foot:parametric}, the action can be written in the form \cite{Kaufman:1987ch}
\begin{equation}
	\Lambda	\doteq	\int \mathrm{d}^4 x \, \mathrm{d}^4x' \, \Psi^* (x)  \mcu{D}(x,x') \Psi(x'),
	\label{def:action_x}
\end{equation}
where $\mcu{D}$ is a Hermitian $[\mcu{D}(x,x')=\mcu{D}^*(x',x)]$ scalar kernel that describes the underlying medium. Varying the action \eq{def:action_x} leads to the following wave equations:
\begin{subequations}\label{eq:basic_ELE_I}
	\begin{align}
		\delta \Psi^*(x): &	\quad	
										0=\int 		\mathrm{d}^4 x' \,	\mcu{D}(x,x') \Psi(x') 		, \\
		\delta \Psi(x): 	&		\quad	
										0 =\int 		\mathrm{d}^4 x' \,	\Psi^*(x') \mcu{D}(x',x) .
	\end{align}
\end{subequations}

For the rest of this work, it will be convenient to describe the wave $\Psi(x)$ also as an abstract vector $\ket{\Psi}$ in the Hilbert space of wave states with inner product \cite{Dodin:2014hw,Littlejohn:1993bd}
\begin{equation}
	\braket{ \Upsilon | \Psi } = \int \mathrm{d}^4 x \, \Upsilon^*(x) \Psi(x).
\end{equation}
In this representation, $\Psi(x)= \braket{x | \Psi}$, where $\ket{x}$ are the eigenstates of the coordinate operator $\hat{x}$ such that $\smash{\braket{ x | \hat{x}^\mu | x' } = x^\mu \braket{ x  | x' } = x^\mu \delta^4(x-x')}$. Let us introduce the momentum (wavevector) operator $\hat{p}$ such that $\smash{\braket{ x | \hat{p}_\mu | x' } = i \pd \delta^4(x-x') / \pd x^\mu }$ in the coordinate representation. Thus, the action \eq{def:action_x} can be rewritten as
\begin{equation}
	\Lambda	=	\bra{\Psi} \oper{D} \ket{\Psi},
	\label{def:action}
\end{equation}
where $\oper{D}$ is the Hermitian \textit{dispersion operator} defined such that $\smash{ \mcu{D}(x,x') = \braket{ x | \oper{D} | x'} }$. Treating $\bra{\Psi}$ and $\ket{\Psi}$ as independent \cite{Dodin:2014hw}, we obtain the following ELEs:
\begin{subequations}\label{eq:basic_ELE}
	\begin{gather}
		\delta \bra{\Psi}: 	\quad		\oper{D} \ket{\Psi} =0, 		\label{eq:ELE_abs_1}		\\
		\delta \ket{\Psi}: 	\quad		\bra{\Psi} \oper{D}  =0,		\label{eq:ELE_abs_2}	
	\end{gather}
\end{subequations}
which can be understood as a generalized vector form of \Eqs{eq:basic_ELE_I}. Specifically, \Eqs{eq:basic_ELE_I} are obtained by projecting \Eqs{eq:ELE_abs_1} and \eq{eq:ELE_abs_2} by $\bra{x}$ and $\ket{x}$, respectively, and using the fact that $\int \mathrm{d} ^4 x \ket{x} \bra{x} = \hat{1}$ is an identity operator.

%%%%%%%%%%%%%%%%%%%%%%%%%%%%%%%%%%%%%%%%%%%%%%
\subsection{Problem outline}

Below, we consider the propagation of a wave $\ket{\Psi}$, called the \textit{probe wave} (PW), in a medium whose parameters are modulated by some other wave, which we call the \textit{modulating wave} (MW). Accordingly, $\mcu{D}(x, x')$ is a rapidly oscillating function. Our goal is to derive a reduced version of \Eqs{eq:basic_ELE} that describes the time-averaged dynamics of the PW. 

We assume that $\oper{D}$ can be decomposed as
\begin{equation}
	\oper{D} 
			=
			\oper{D}_0
			+  \oper{D}_{\rm osc},
	\label{def:D}
\end{equation}
where $\smash{\oper{D}_0}$ represents the effect of the unperturbed background medium and $\smash{\oper{D}_{\rm osc}}$ represents a weak perturbation caused by the MW. Additionally, we assume
\begin{equation}
	\oper{D}_{\mathrm{osc}}
		= \sum_{n=1}^\infty	\sigma^n \oper{D}_n,
	\label{def:Dosc}
\end{equation}
where $\sigma \ll 1$ is some linear measure of the MW amplitude \cite{foot:sigma} and $\smash{\oper{D}_n}$ are Hermitian. Finally, we require the MW (but not necessarily the PW) to satisfy the standard assumptions of geometrical optics (GO). This means that the MW frequency $\Omega$ and wave vector $\vec{K}$ must be large compared to the inverse temporal and spatial scales at which the envelope evolves. In a homogeneous medium, those scales would be simply the MW envelope duration $\tau_{\rm mw}$ and the MW envelope length $\ell_{\rm mw}$. More generally, one also has the scales $\tau_{\rm bg}$ and $\ell_{\rm bg}$ that characterize the background temporal and spatial inhomogeneities, correspondingly. Thus, the applicability of our theory relies on the smallness of the following parameter:
\begin{equation}
	\epsilon_{\rm mw} \doteq 
			\mathrm{max} \left\lbrace    
			\frac{1}{\Omega \tau}	, 
			\frac{1}{ |\vec{K}| \ell } 
			\right\rbrace \ll 1,
	\label{eq:assumption}
\end{equation}
where $\tau \doteq \text{min}\,\{\tau_{\rm bg}, \tau_{\rm mw}\}$ and $\ell \doteq \text{min}\,\{\ell_{\rm bg}, \ell_{\rm mw}\}$. A more rigorous definition of the GO regime that covers also waves near natural resonances is somewhat subtle, so it is not discussed here. For details, the reader is referred, \eg to \Ref{Tracy:2014to}.

%%%%%%%%%%%%%%%%%%%%%%%%%%%%%%%%%%%%%%%%%%%%%%
%%%%%%%%%%%%%%%%%%%%%%%%%%%%%%%%%%%%%%%%%%%%%%

\section{General theory}
\label{sec:gen_theory}

\begin{figure}
	\includegraphics[scale=.40]{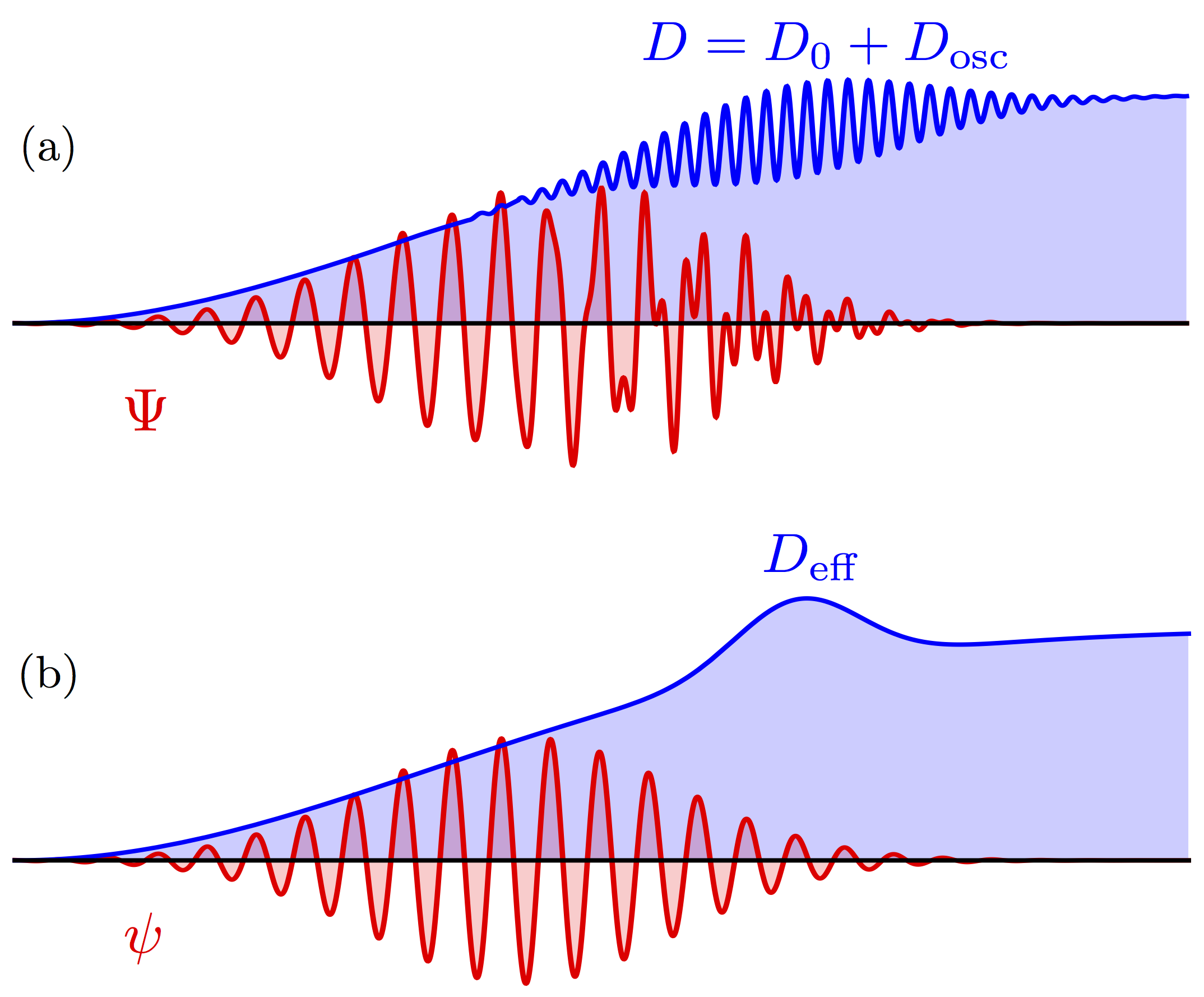}
	\caption{One-dimensional schematic of a PW (in red) propagating in a medium with a given dispersion operator affected by some MW. (a) Dynamics in the original variables. (b) Dynamics in the OC representation, in which the oscillations at the MW phase and its harmonics are removed.}
	\label{fig:transformation}
\end{figure}

The oscillating terms in the dispersion operator will be eliminated by introducing an appropriate variable transformation on the PW. Specifically, let $\smash{\ket{\Psi} = \oper{U} \ket{\psi}}$. Then, \Eq{def:action} transforms to
\begin{equation}
	\Lambda	= \bra{\psi} \oper{D}_{\rm eff} \ket{\psi},
	\label{def:act_eff}
\end{equation}
where $\oper{D}_{\rm eff}$ is the effective dispersion operator
\begin{equation}
	\oper{D}_{\rm eff} \doteq 
			\oper{U}^\dag 
			\oper{D}
			\oper{U}.
	\label{def:D_eff}
\end{equation}
Below, we search for a transformation $\smash{\oper{U}}$ such that, unlike $\smash{\oper{D}}$, the operator $\smash{\oper{D}_{\rm eff}}$ contains no dependence on the MW phase. The corresponding $\ket{\psi}$ is then understood as the OC state of the PW in a modulated medium. A schematic of the transformation is shown in \Fig{fig:transformation}.

%%%%%%%%%%%%%%%%%%%%%%%%%%%%%%%%%%%%%%%%%%%%%%
\subsection{Near-identity unitary transformation}

For convenience, we require that $\smash{\oper{U}}$ be unitary so that $\braket{\Psi | \Psi } = \braket{ \psi | \psi }$. Then, it can be represented as
\begin{equation}
	\oper{U} = \exp (  i \oper{T}	),
	\label{eq:U}
\end{equation}
where $\smash{\oper{T}}$ is a Hermitian operator called the \textit{generator} of the unitary transformation $\smash{\oper{U}}$. In light of \Eq{def:Dosc}, we search for $\smash{\oper{T}}$ and $\smash{\oper{D}_{\rm eff}}$ using the standard perturbation approach based on Lie transforms \cite{Cary:1981tw,Cary:1983dy}. Specifically, we consider the operators as power series in $\sigma$ so that
\begin{equation}
	\oper{T}= \sum_{n=1}^\infty \sigma^n \oper{T}_n, 	\quad
	\oper{D}_{\rm eff} = \sum_{n=0}^\infty \sigma^n \oper{D}_{\mathrm{eff}, n} ,
	\label{eq:Deff_expansion}
\end{equation} 
where $\smash{\oper{T}_n}$ and $\smash{\oper{D}_{\mathrm{eff}, n}}$ are Hermitian. We substitute \Eqs{def:D}, \eq{def:Dosc}, \eq{eq:U}, and \eq{eq:Deff_expansion} into \Eq{def:D_eff}. Collecting terms by equal powers in the parameter $\sigma$, we obtain the following set of equations \cite{Gramespacher:1996kz}:
\begin{subequations}\label{eq:order}
	\begin{align}
		\oper{D}_{\mathrm{eff}, 0} = 	&	\, \oper{D}_0, 				\label{eq:order0} 	\\
		\oper{D}_{\mathrm{eff}, 1} =	&	\, \oper{D}_{1} 
														+ i [\oper{D}_0, \oper{T}_1  ], 		
														\label{eq:order1}		\\
		\oper{D}_{\mathrm{eff}, 2} =	& \, \oper{D}_2 
														+ i [ \oper{D}_0,\oper{T}_2  ]
														+ \oper{C}_2, 
														\label{eq:order2}	
	\end{align}
\end{subequations}
where $\smash{ \oper{C}_2 \doteq i [\oper{D}_1, \oper{T}_1 ] - (1/2)  \boldsymbol{[} [\oper{D}_0 , \oper{T}_1 ], \oper{T}_1 \boldsymbol{]} }$
and so on. Here the brackets denote commutators. We require that $\smash{\oper{D}_{\mathrm{eff}, n}}$ contains no high-frequency modulations, so we let
\begin{subequations} \label{eq:Deffn}
	\begin{align}
		\oper{D}_{\mathrm{eff}, 1} =	& \, \avg{ \oper{D}_{1} } , 	\label{eq:Deff1} \\
		\oper{D}_{\mathrm{eff}, 2} =	& \, \avg{ \oper{D}_2 } + \avg{ \oper{C}_2} , \label{eq:Deff2}
	\end{align}
\end{subequations}
where `$\smash{\avg{...}}$' is a time average over a modulation period. Then, subtracting \Eqs{eq:Deffn} from \Eqs{eq:order}, we obtain
\begin{subequations} \label{eq:Tn}
	\begin{align}
		-i [\oper{D}_0, \oper{T}_1 ]  = & \, \oper{D}_1 -	\avg{ \oper{D}_1 }  , 	\label{eq:T1} \\
		-i [\oper{D}_0, \oper{T}_2 ]  = & \, \oper{D}_2 -	\avg{ \oper{D}_2 } 
															+	\oper{C}_2 - \avg{ \oper{C}_2	}. \label{eq:T2}
	\end{align}
\end{subequations}
As usual, this procedure can be iterated to higher orders in $\sigma$. However, for the sake of conciseness, we shall only calculate $\oper{D}_{\rm eff}$ up to $\mcu{O}(\sigma^2)$ in this work. Below, we demonstrate how to solve \Eqs{eq:Tn} for $\oper{T}_1$ and $\oper{T}_2$.

%%%%%%%%%%%%%%%%%%%%%%%%%%%%%%%%%%%%%%%%%%
\subsection{$\boldsymbol{\mcu{D}_{\rm eff}}$ within the leading-order approximation}
\label{sec:Deff}

In order to explicitly obtain $\smash{\oper{D}_{\rm eff}}$ and $\oper{T}_n$, let us consider \Eqs{eq:order}-\eq{eq:Tn} in the Weyl representation. (Readers who are not familiar with the Weyl calculus are encouraged to read \App{app:Weyl} before continuing further.) For $n=0$, the Weyl transformation of \Eq{eq:order0} leads to
\begin{equation}
	D_{\mathrm{eff}, 0}(x,p) =	D_{0}(x,p) ,
	\label{eq:trans_0}
\end{equation}
where $D_n( x,p)$ and $D_{\mathrm{eff},n}( x,p)$ are the Weyl symbols \eq{def:weyl_symbol} of the operators $\smash{\oper{D}_n}$ and $\smash{\oper{D}_{\mathrm{eff},n}}$, respectively. For $n=1$, the Weyl transformation of \Eq{eq:order1} gives
\begin{equation}
	D_{\mathrm{eff}, 1} =	D_{1} - \moysin{ D_0 , T_1} ,
	\label{eq:trans_1}
\end{equation}
where `$\moysin{\cdot,\cdot }$' is the Moyal sine bracket \eq{eq:sine_bracket} and $T_n(x,p)$ are the Weyl symbols of $\smash{\oper{T}_n}$. It is to be noted that $D_n$, $D_{\mathrm{eff},n}$, and $T_n$ are real functions of the eight-dimensional phase space because the corresponding operators are Hermitian.

Since $D_1$ is a linear measure of the MW field, we adopt
\begin{equation}
	D_1 (x,p)=  \mathrm{Re}	[ \mc{D}_1(x,p) e^{i\Theta(x) }  ],
	\label{eq:D_1}
\end{equation}
where the real function $\Theta(x)$ is the MW phase and $\smash{\mc{D}_1(x,p)}$ is the Weyl symbol characterizing the slowly-varying MW envelope \cite{foot:many_waves,foot:slowly}. The gradients of the phase 
\begin{equation}
	\Omega( x) \doteq -\pd_t \Theta , \quad \vec{K}( x) \doteq \del \Theta ,
	\label{eq:MW_Omega_K}
\end{equation}
determine the MW local frequency and wavevector, respectively. We introduce the MW four-wavevector $K_\mu( x)\doteq -\pd_\mu \Theta = (\Omega, -\vec{K})$, which is considered a slow function. [Accordingly, the contravariant representation of the MW four-wavevector is $K^\mu(x) = (\Omega, \vec{K})$.]

Since $D_1$ is quasi-periodic \cite{foot:quasi}, we have $\avg{D_1}=0$. Following \Eq{eq:Deff1}, then $D_{\mathrm{eff}, 1} =0$, which also gives
\begin{equation}
	\moysin{ D_0 , T_1 } =  D_{1} .
	\label{eq:trans_1_zero}
\end{equation}
Let us search for $T_1$ in the polar representation:
\begin{equation}
	T_1 =  \mathrm{Re}	[ \mc{T}_1( x,p) e^{i\Theta( x)}  ] ,
	\label{eq:T_1}
\end{equation}
where $\mc{T}_1(x,p)$ is to be determined. Substituting \Eqs{eq:D_1} and \eq{eq:T_1} into \Eq{eq:trans_1_zero} and equating terms with the same phase, we obtain (\App{app:auxiliary})
\begin{align}
	\mc{D}_1(x,p)e^{i\Theta(x)}
	 = 		&		\,
	 					\moysin{	D_0	 , \mc{T}_1		e^{i\Theta} }	 
						\notag \\
	= 			&		\, \mc{T}_1
						\moysin{	D_0	 , 	e^{i\Theta} }
						+ \mcu{O}(\epsilon_{\rm mw})
						\notag \\
	= 			&		\, -i\mc{T}_1
						(	D_0	 \star  	e^{i\Theta} - e^{i\Theta}	\star	D_0	) 		\, 		
						+ \mcu{O}(\epsilon_{\rm mw})
						\notag \\
	=			& 		\, -i
						\left[ 	D_0 ( x,p+K/2 )	- 	D_0 ( x,p-K/2)		\right]
						\notag \\
				&		~~~	\times \mc{T}_1(x,p) e^{i\Theta(x)} 
						+ \mcu{O}(\epsilon_{\rm mw}),
	\label{eq:aux1}
\end{align}
where `$\star$' is the Moyal product \eq{def:Moyal} and $\mc{T}_1$ is pulled out of the sine bracket because it is a slowly-varying function. Solving for $\mc{T}_1$, we obtain
\begin{equation}
	\mc{T}_1( x,p) = 
		 \frac{ i \mc{D}_1( x,p)}{D_0( x,p+K/2)-D_0( x,p-K/2)} +\mcu{O}(\epsilon_{\rm mw}).
	\label{eq:G}
\end{equation}

Now let us calculate $D_{ \mathrm{eff},2}$. From \Eq{eq:order1}, we have $\smash{ [ \oper{D}_0, \oper{T}_1]=i\oper{D}_1}$, so $\smash{ \oper{C}_2 = - (i/2) [\oper{T}_1, \oper{D}_1 ] }$. Then, by applying the Weyl transform to \Eq{eq:order2}, we obtain
\begin{equation}
	D_{\mathrm{eff}, 2} 	=	D_{2} - \moysin{  D_0 , T_2 } +C_2,
	\label{eq:trans_2}
\end{equation}
where $C_2(x,p) = (1/2) \moysin{  T_1 , D_1}$. After substituting $D_1$ and $T_1$, the Weyl symbol $C_2$ is found to be (\App{app:auxiliary})
\begin{align}
	C_2(x,p) =	&  	-	\frac{1}{4}		
								\sum_{n = \pm 1}
									\frac{| \mc{D}_1( x,p+nK/2 )|^2 }{D_0( x,p+nK)-D_0( x,p)} 
								\notag \\
		 				&		+\mathrm{Re} [ \mc{C}_2( x,p) e^{i2\Theta(x)} 	] 
		 + \mcu{O}(\epsilon_{\rm mw}),
		 \label{eq:C2_aux}
\end{align}
where $\mc{C}_2(x,p)$ is a slowly-varying function whose explicit expression will not be needed for our purposes.

Following \Eqs{eq:Deff2} and \eq{eq:trans_2}, we let $D_{\mathrm{eff}, 2} = \avg{D_2} + \avg{C_2}$. Then, the symbol $T_2(x,p)$ satisfies 
\begin{equation}
	\moysin{D_0, T_2 }
		= D_2 - \avg{D_2}
		+  \mathrm{Re} ( \mc{C}_2 e^{i2\Theta }	) .
	\label{eq:trans_3}
\end{equation}
We then repeat the procedure shown in \Eqs{eq:T_1}-\eqref{eq:G} to obtain $T_2$ that satisfies \Eq{eq:trans_3}. Finally, after collecting the previously obtained results of this section, the effective dispersion symbol is found to be
 \begin{align}
	D_{\rm eff}(x,p) 
			= 	&	\,	D_0 (x,p) 
						+ \sigma^2 \avg{D_2(x,p) } \notag \\
				&	- \frac{\sigma^2}{4} \sum_{n = \pm 1} 
					\frac{|\mc{D}_1(x,p+nK /2)|^2}{ D_0 (x,p+nK)- D_0 (x,p)} 	\notag\\
				&	+ \mcu{O}(\epsilon_{\rm mw},\sigma^4).
	\label{eq:Deff_final}
\end{align}
The leading-order correction to $D_{\rm eff}(x,p)$, that scales as $\epsilon_{\rm mw}^0$, can be only of the fourth power of $\sigma$. This occurs because the third and other odd powers of the MW field have zero average and thus cannot contribute to the effective dispersion symbol $D_{\rm eff}(x,p)$ that governs the $\Theta$-averaged motion.

The Weyl symbol $D_{\rm eff}(x,p)$ in \Eq{eq:Deff_final} constitutes one of the main results of this work. It determines the asymptotic form of the effective dispersion operator that governs the dynamics of the PW averaged over the MW oscillations at small enough GO parameter $\epsilon_{\rm mw}$ [\Eq{eq:assumption}] and small enough MW amplitude $\sigma$. The actual operator $\smash{\oper{D}_{\rm eff}}$ can be obtained from the symbol \eq{eq:Deff_final} using the inverse Weyl transform \eq{eq:weyl_inverse}. Alternatively, one can find its coordinate representation $\mcu{D}_{\rm eff}(x, x')$ using \Eq{eq:weyl_x_rep}.

%%%%%%%%%%%%%%  PONDEROMOTIVE DYNAMICS %%5%%%%%%%%%%%%
%%%%%%%%%%%%%%%%%%%%%%%%%%%%%%%%%%%%%%%%%%%%%%
\section{Ponderomotive dynamics}
\label{sec:pondero}

With the effective dispersion operator $\smash{\oper{D}_{\rm eff}}$, we can describe the time-averaged dynamics of the PW using
\begin{gather}
	\oper{D}_{\rm eff} \ket{\psi} = 0.
\end{gather}
Alternatively, we can apply the variational approach and study the action \eq{def:act_eff} in the \textit{phase-space representation}. Following \Refs{Kaufman:1987ch,Tracy:2014to}, the action is written as
\begin{equation}
	\Lambda= \int \mathrm{d}^4 x \, \mathrm{d}^4p \,
					 D_{\rm eff}(x,p)W(x,p) ,
	\label{eq:action_eff}
\end{equation}
where $W(x,p)$ is the Wigner function \cite{Wigner:1932cz} corresponding to the OC state $\ket{\psi}$; namely,
\begin{equation}
	W(x,p) \doteq \int \frac{\mathrm{d}^4 s}{(2\pi)^4} \, 
					e^{i  p \cdot s } 
					\braket{x+s/2 | \psi} 
					\braket{\psi | x-s/2} .
	\label{def:wigner_function}
\end{equation}

The variational approach is particularly convenient for deriving approximate models of wave dynamics \cite{Friedland:1987be,Cook:1993vt,Tracy:2003bl,Whitham:2011kb,Ruiz:2015hq,Dodin:2014dv,Dodin:2014iz,Ruiz:2015dv,Balakin:2016kt}. For illustration purposes, here we focus on the OC dynamics of PWs in the eikonal approximation. Specifically, we proceed as follows.

%%%%%%%%%%%%%%%%%%%%%%%%%%%%%%%%%%%%%%%%%%%%%%
\subsection{Eikonal approximation}

Let us consider the complex function $\psi \doteq \braket{x|\psi}$ in the following polar representation
\begin{equation}
	\braket{x | \psi}  	
					=	\psi(x) 
					=\sqrt{ \mc{I}_0(x) } \, e^{i\theta(x)},
	\label{eq:semiclassical}
\end{equation}
where $\mc{I}_0(x)$ and $\theta(x)$ are real functions. We assume that the phase $\theta$ is fast compared to the slowly-varying function $\mc{I}_0$. We also assume 
\begin{equation}
	\epsilon_{\rm pw} \doteq 
			\mathrm{max} \left\lbrace    
			\frac{1}{\omega \tau}	, 
			\frac{1}{ |\vec{k}| \ell } 
			\right\rbrace \ll 1,
	\label{eq:assumption2}
\end{equation}
where 
\begin{equation}
	\omega(x) \doteq - \pd_t \theta, \quad \vec{k}(x) \doteq \del \theta
\end{equation}
are the local PW frequency and the wavevector, respectively. In other words, we consider that the characteristic scale lengths of the inhomogeneities of the background medium and of the MW envelope are large with respect to the wavelength of the PW. For simplicity, we combine the small parameters \eq{eq:assumption} and \eq{eq:assumption2} into a single parameter
\begin{equation}
	\epsilon \doteq \mathrm{max} \{ \epsilon_{\rm mw}, \epsilon_{\rm pw} \} \ll 1.
\end{equation}
In particular, note that in order to apply the standard GO approximation to the original problem, the PW parameters must satisfy $\Omega/\omega \ll 1$ and $|\vec{K}|/ |\vec{k}| \ll 1$; \eg the PW wavelength must be smaller than the MW wavelength. However, after the transformation, the MW oscillations are eliminated (see \Fig{fig:transformation}), so the PW parameters must satisfy the less restrictive condition \eq{eq:assumption2}.

Since $\psi$ is assumed quasi-monochromatic, the Wigner function \eq{def:wigner_function} is then, to the lowest order in $\epsilon$ \cite{Kaufman:1987ch},
\begin{equation}
	W(x,p)= \mc{I}_0 (x)  \delta^4( p - k)	+ \mcu{O}(\epsilon),
	\label{eq:T_wigner}
\end{equation}
where $k_\mu(x) \doteq -\pd_\mu \theta= (\omega, -\vec{k})$. Substituting \Eq{eq:T_wigner} into \Eq{eq:action_eff} leads to the following action:
\begin{equation}
	\Lambda =  \int \mathrm{d}^4 x   \, \mc{I}_0(x)  D_{\rm eff}(x,k) .
	\label{eq:action_eikonal}
\end{equation}

The action \eq{eq:action_eikonal} has the form of Whitham's action, where
\begin{equation}
	\mc{I} \doteq \mc{I}_0 \pd_\omega D_{\rm eff} (x, k) 
	\label{eq:act_den}
\end{equation}
serves as the wave action density \cite{Whitham:2011kb}. [From now on, $k(x)=-\pd \theta$.] Treating $\mc{I}_0$ and $\theta$ as independent variables yields the following ELEs:
\begin{subequations}	\label{eq:ELE_fluid_eff}
	\begin{align}
		\delta \theta: 	\quad 	&	\pd_t \mc{I}  
												+ \del \cdot (  \mc{I} \vec{v} )=0,	\label{eq:act}	\\
		\delta \mc{I}_0: \quad 	&	D_{\rm eff}(x, k)=0,									\label{eq:hj}
	\end{align} 
\end{subequations}
where the flow velocity $\vec{v}$ is given by
\begin{equation}
	\vec{v}(t,\vec{x}) \doteq - \frac{ \pd_\vec{\vec{k}} D_{\rm eff} }{ \pd_\omega D_{\rm eff} }.
	\label{eq:v}
\end{equation}
Equation \eq{eq:act} represents the action conservation theorem, and \Eq{eq:hj} is the local wave dispersion relation.

%%%%%%%%%%%%%%%%%%%%%%%%%%%%%%%%%%%%%%%%%%%%%%
\subsection{Hayes's representation}
\label{sec:Hayes}

Equation \eq{eq:hj} can be used to express the PW frequency $\omega$ as some function $H_{\rm eff}(t, \vec{x}, \del \theta)$:
\begin{equation}
	\omega = H_{\rm eff}(t,\vec{x}, \del \theta) .
	\label{eq:omega}
\end{equation}
This determines a \textit{dispersion manifold} \cite{Tracy:2014to,foot:multiple}. The function $H_{\rm eff}$ can be represented as follows:
\begin{equation}
	H_{\rm eff}(t,\vec{x},\vec{k}) \doteq 
			H_0(t, \vec{x},\vec{k})
			+\sigma^2 \Phi (t,\vec{x},\vec{k})   ,
	\label{eq:Heff}
\end{equation}
where higher powers of $\sigma$ are neglected, like in the previous section. (Henceforth, the small parameter $\sigma$ will be omitted for clarity.) Here $H_0(t, \vec{x},\vec{k}) $ is the \textit{unperturbed frequency} of the PW, so it satisfies $D_0(x,k_*)=0$, where 
\begin{equation}
 	k_*^\mu(t,\vec{x},\vec{k}) \doteq \boldsymbol{(} H_0(t,\vec{x},\vec{k}), \vec{k} \boldsymbol{)}
\end{equation}
is the unperturbed PW four-wavevector. The function $\Phi(t, \vec{x},\vec{k})$ can be understood as the PW \textit{ponderomotive frequency shift}. When multiplied by $\hbar$, $\Phi$ is also understood as the ponderomotive energy or ponderomotive potential; that said, one may want to restrict usage of the term ``potential'' to cases when $\Phi$ is independent of $\vec{k}$.

Using \Eqs{eq:Deff_final} and \eq{eq:hj} together with the Taylor expansion
\begin{equation}
	D_{\rm eff} (x,k) \approx  
					D_{\rm eff} (x,k_*) 
					+\pd_\omega D_{\rm eff}(x,k_*) [ \omega - H_0(t,\vec{x},\vec{k})] ,
	\label{eq:D_Taylor}
\end{equation}
we obtain an explicit expression for $\Phi$, which is
\begin{multline}
	\Phi (t,\vec{x},\vec{k})	=	\bigg[
			- \frac{	\avg{D_2(x,k) }	}{\pd_\omega D_0( x, k ) }	
			+ \frac{1}{4 \pd_\omega D_0( x, k) } \\
				\times 
				\sum_{n = \pm 1}
						\frac{ | \mc{D}_1( x, k + n K/2 )|^2 } 
						{D_0(x, k	+n K ) 
								- D_0(x, k  ) } \bigg]_{k=k_*}.
	\label{eq:pondero_potential}
\end{multline}

Hence, we can rewrite the action \eq{eq:action_eikonal} in the Hayes's form \cite{Hayes:1973dt}; namely,
\begin{equation}
	\Lambda \simeq - \int \mathrm{d}^4 x \, 
								\mc{I} \left[ \pd_t \theta + H_{\rm eff}(t,\vec{x}, \del \theta)   \right] .
	\label{eq:action_Hayes}
\end{equation} 
In this case, the corresponding ELE's are
\begin{subequations}\label{eq:Hayes}
	\begin{align}
			\delta \theta: 	\quad 	&	\pd_t  \mc{I}  + \del \cdot ( \mc{I} \vec{u} )=0	,		\label{eq:act_eff}	\\
			\delta \mc{I}: 	\quad 	&	\omega = H_{\rm eff}(t,\vec{x}, \vec{k}),	\label{eq:hj_eff}
	\end{align} 
\end{subequations}
where $\vec{u}$ is the effective PW group velocity,
\begin{equation}
	\vec{u}(t,\vec{x})	\doteq	 \pd_\vec{k} H_{\rm eff}(t,\vec{x},\vec{k}).
\end{equation}
Equation \eq{eq:hj_eff} is a Hamilton--Jacobi equation representing the local wave dispersion. Note that, on solutions of \Eq{eq:hj_eff}, $\vec{u}(t,\vec{x})$ is the same as $\vec{v}(t,\vec{x})$ defined in \Eq{eq:v}, so \Eqs{eq:Hayes} are consistent with \Eqs{eq:ELE_fluid_eff}.

Another comment is the following. When $ | K | \ll | k_* | $, the effective Hamiltonian can be approximated to
\begin{align}
	H_{\rm eff} & (t,\vec{x},  \vec{k}) \simeq 
			H_0(t, \vec{x},\vec{k}) + \bigg[ - \frac{	\avg{ D_2(x,k) }	}{\pd_\omega D_0( x, k ) }	 
				\notag \\
			& +  \frac{\sigma^2}{4 \pd_\omega D_0( x, k)}  
					 K^\mu \frac{\pd}{\pd k^\mu }
						\left( \frac{ | \mc{D}_1( x, k )|^2 } {K^\nu \pd_{k^\nu} D_0(x,k)  } \right)
						\bigg]_{k=k_*},
\label{eq:H_dodin}
\end{align}
where $K^\mu \pd_{k^\mu}	\doteq	\Omega \pd_\omega + \vec{K} \cdot \pd_{\vec{k}}$. When $D(x,p)$ in \Eq{def:D} is of the Hayes's form $[D(x,p) = p_0 - H(t, \vec{x}, \vec{p} ) ]$, \Eq{eq:H_dodin} recovers the same expression for $H_{\rm eff}$ that was previously reported in \Ref{Dodin:2014iz}.

%%%%%%%%%%%%%%%%%%%%%%%%%%%%%%%%%%%%%%%%%%%%
\subsection{Point-particle model and ray equations}
\label{sec:rays}

The ray equations corresponding to \Eqs{eq:Hayes} can be obtained by assuming the point-particle limit. Specifically, let us adopt the ansatz
\begin{equation}
	\mc{I}(t,\vec{x})= \delta^3 \textbf{(} \vec{x}-\vec{X}(t)  \textbf{)},
\end{equation}
where $\vec{X}(t)$ is the location of the wave packet. As in \Ref{Ruiz:2015bz}, integrating the action \eq{eq:action_Hayes} in space yields the canonical phase-space action of a point-particle; namely,
\begin{equation}	\label{eq:lagr_point}
	\Lambda	= \int \mathrm{d} t \, 
						[ \vec{P} \cdot \dot{\vec{X}} - H_{\rm eff}(t,\vec{X},\vec{P}) ] ,
\end{equation}
where $\vec{P}(t) \doteq \del \theta \textbf{(} t, \vec{X}(t) \textbf{)} $. Here, $\vec{X}(t)$ and $\vec{P}(t)$ serve as canonical variables. The corresponding ELEs are
\begin{subequations}\label{eq:ray}
	\begin{align}
		\delta \vec{P} :  &	\quad  \dot{\vec{X}} 	= 	\pd_\vec{P} H_{\rm eff}(t,\vec{X},\vec{P} ), \\
		\delta \vec{X} :  &	\quad  \dot{\vec{P}}	=  	-  \pd_\vec{X} H_{\rm eff}(t,\vec{X},\vec{P}).
	\end{align}
\end{subequations}
Equations \eq{eq:ray} describe the ponderomotive dynamics of PW rays. These equations include the time-averaged refraction of a PW caused by the MW oscillations. The ponderomotive dynamics of charged particles is subsumed here as a special case. (Also note that, since $H_{\rm eff}$ is generally not separable into a kinetic energy and a potential energy, the dynamics governed by \Eqs{eq:ray} may be quite complicated and perhaps counter-intuitive \cite{my:mneg, my:trieste08}.) Some examples are discussed below. 

%%%%%%%%%%%%%%%%%%%%%%%%%%%%%%%%%%%%%%%%%%%%%%
%%%%%%%%%%%%%%%%%%%%%%%%%%%%%%%%%%%%%%%%%%%%%%
\section{Discussion and examples}
\label{sec:applications}

%%%%%%%%%%%%%%%%%%%%%%%%%%%%%%%%%%%%%%%%
\subsection{Example 1: Schr\"odinger particle in an electrostatic field}
\label{sec:schro}

We consider a nonrelativistic particle interacting in a modulated electrostatic potential. The particle dynamics can be described using the Schr\"odinger equation
\begin{equation}
	i \pd_t \Psi = \left[ -\del^2 / 2m +  q V(x) \right] \Psi,
	\label{eq:schro_main}
\end{equation}
where $m$ and $q$ are the particle mass and charge, the electrostatic potential $V(x)=\mathrm{Re}	\left[ V_{\rm c} (x) e^{i\Theta(x) }\right]$ is assumed small, $\Theta(x)$ is a real fast phase, and $V_{\rm c} (x)$ is a complex function describing the slowly-varying potential envelope. In this case, the dispersion operator is
\begin{equation}
	\oper{D} \doteq \hat{p}_0- \hat{\vec{p}}^2/2m-  qV(\hat{x}).
\end{equation}
The corresponding Weyl symbols are (\App{app:Weyl})
\begin{equation}
	D_0(p ) 			=  p_0 - \vec{p}^2/2m, \quad
	D_{\rm osc}(x) 	= - qV(x)	
	\label{eq:schro_D}.
\end{equation}

\begin{figure}
	\centering
	\includegraphics[scale=0.6]{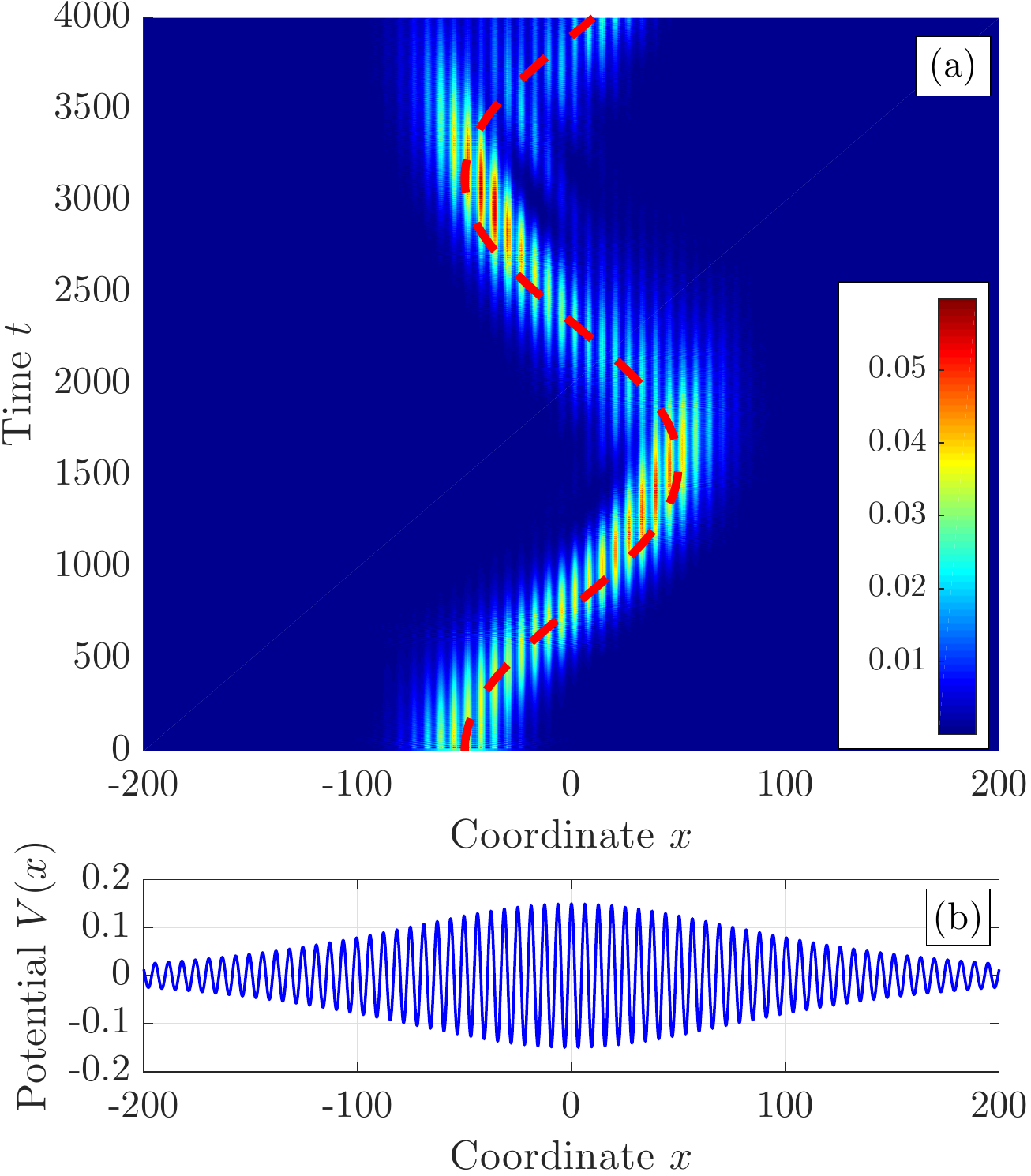}
	\caption{
(a) Comparison of the simulation results obtained by numerically integrating the full-wave \Eq{eq:schro_main} (solid fill) and the ray-tracing \Eqs{eq:ray} with $H_{\rm eff}$ taken from \Eq{eq:schro_Heff} (dashed) for a stationary MW. The initial wave packet is $\Psi_0(x)=(2 \pi \eta^2 )^{-1/4} \exp \left[ - (x-\mu)^2/(4 \eta^2) \right]$, where $\mu=-50$ and $\eta = 15$, and it is normalized such that $\smash{\int \mathrm{d}x \, \Psi_0^2(x)=1}$. (The simulation is one-dimensional, and $x$ denotes the spatial coordinate, unlike in the main text, where $x$ denotes the spacetime coordinate.) The initial conditions for the ray trajectory are $X(0)=-50$ and $\smash{P(0)=0}$. (b) MW profile of the form $V(x) = 0.15~ \mathrm{sech}(x/80) \cos (x)$. Natural units are used such that $m=1$, $q=1$, $\hbar =1$, and $K=1$. At later times (not shown), diffraction effects become important, so the eikonal theory becomes inapplicable.}
	\label{fig:trapping}
\end{figure}

The symbol $D_{\rm eff}$ is calculated using \Eq{eq:Deff_final}. Note that $ \smash{D_1=  -\mathrm{Re} ( q V_{\rm c}  e^{i\Theta} ) }$ so $\mc{D}_1 =-q V_c$ and $D_n=0$ for $n\geq 2$. Substituting into \Eq{eq:Deff_final}, we obtain
 \begin{align}
	&	D_{\rm eff}		(x,p) \notag  \\
	&	 	=  D_0 (p)
				- \frac{1}{4} \sum_{n = \pm 1}
				\frac{|\mc{D}_1(x)|^2}{ D_0 (p+nK)- D_0 (p)}		\notag \\
	&		=	  p_0 - \frac{\vec{p}^2}{2m}  
				-\sum_{n = \pm 1}
					\frac{q^2|V_c(x)|^2/4}{ \left( p_0+n \Omega - \frac{(\vec{p}+n \vec{K})^2}{2m}	\right)
							- \left( p_0-  \frac{\vec{p}^2}{2m} \right)}		
				\notag \\
	&	 =  p_0 - \frac{\vec{p}^2}{2m}
				- \frac{ q^2 |\vec{K} V_c|^2 /m}
				{ 4( \Omega -  \vec{p} \cdot \vec{K}/m)^2	- ( \vec{K}^2/m )^2}	.
	\label{eq:schro_Deff}
\end{align}
Inserting \Eq{eq:schro_Deff} into \Eq{eq:action_eikonal} leads to the action in the Hayes form \eq{eq:action_Hayes}, where the effective Hamiltonian is
\begin{equation} 
H_{\rm eff}(t,\vec{x},\vec{k}) =
				\frac{\vec{k}^2}{2m} + \frac{ q^2|\vec{K} V_c|^2 /m}{ 4(\Omega - \vec{k}\cdot \vec{K}/m)^2 - (\vec{K}^2 /m)^2 } .
\end{equation}
In the fluid description of the particle wave packet, the corresponding ELEs are given by \Eqs{eq:Hayes}.  The corresponding ray equations are obtained from the point-particle Lagrangian \eq{eq:lagr_point}. When introducing the missing $\hbar$ factors, the effective Hamiltonian becomes
\begin{equation}
	H_{\rm eff}(t,\vec{X},\vec{P}) 
			=	\frac{\vec{P}^2 }{2m} +
				\frac{ q^2|\vec{K} V_c|^2 /m}
					{ 4(\Omega - \vec{P} \cdot \vec{K}/m)^2 - ( \hbar \vec{K}^2 /m)^2 }.
	\label{eq:schro_Heff}
\end{equation}

In contrast with the classical ponderomotive Hamiltonian \cite{ref:dewar72c,Cary:1977ud,Dodin:2014dv}
\begin{equation}
	H_{\rm eff,cl}(t,\vec{X},\vec{P})=  
			\frac{\vec{P}^2}{2m}
			+  \frac{ q^2|\vec{K} V_c|^2}{4m(\Omega - \vec{P} \cdot \vec{K}/m)^2},
	\label{eq:schro_H_GM}
\end{equation}
which is recovered from \Eq{eq:schro_Heff} at small enough $\vec{K}$, \Eq{eq:schro_Heff} predicts that the ponderomotive force can be attractive. This is seen from the fact that, when the second term in the denominator in \Eq{eq:schro_Heff} dominates, the ponderomotive energy becomes an effective negative potential (\ie does not depend on $\vec{P}$). This effect, which is similar to that reported in \Refs{Gilary:2003bj,Rahav:2003it}, was confirmed in our numerical simulations, whose results are shown in \Fig{fig:trapping}. Note that the ray trajectories generated by $H_{\rm eff}$ accurately match the motion of the wave packet's center. 

Also note that, at $\Omega = 0$, the ponderomotive energy is resonant at $ 2 \vec{K} \cdot \vec{P} = \pm \hbar \vec{K}^2 $. This relation can be written as  $\lambda_{\rm dB} = 2 d \cos\zeta$, where $\lambda_{\rm dB}$ is the particle de Broigle wavelength, $d\doteq 2\pi / K$ is the characteristic length of the lattice, and $\zeta$ is the angle between the $\vec{K}$ and $\vec{P}$ vectors. One may recognize this as the Bragg resonance. In other words, our wave theory presents Bragg scattering as a variation of the ponderomotive effect. One can also identify a parallel between \Eq{eq:schro_Heff} and the linear susceptibility of quantum plasma \cite{Lifshitz:1981ui,book:Haas,Daligault:2014ge}. This will be further discussed in \Sec{sec:mod}.

%%%%%%%%%%%%%%%%%%%%%%%%%%%%%%%%%%%%
\subsection{Example 2: Ponderomotive dynamics of a relativistic spinless particle}

In this section we calculate the ponderomotive Hamiltonian of a relativistic spinless particle interacting with a slowly-varying background EM field and a high-frequency EM modulation \cite{foot:pondero_spin}. The particle dynamics can be described using the Klein--Gordon equation
\begin{equation}
	\left[ (i \pd_t - q V )^2 - (-i \del - q \vec{A})^2 - m^2  \right] \Psi = 0,
	\label{eq:KG_main}
\end{equation}
where $V(x)$ and $\vec{A}(x)$ are the scalar and vector potentials, respectively. Let $A^\mu(x) \doteq(V,\vec{A})$ be the associated four-potential, which can be written as
\begin{equation}
	A^\mu (x) \doteq A^\mu_{\rm bg} (x) +  A^\mu_{\rm osc}(x).
\end{equation}
Here $A^\mu_{\rm bg} (x)$ is the four-potential describing the background EM field, and $\smash{A^\mu_{\rm osc} (x) \doteq \mathrm{Re} [ A^\mu_c (x) e^{i\Theta(x) } ]}$ is the four-potential of the modulated EM wave with small amplitude. As before, $\Theta(x)$ is a real fast phase, and $A^\mu_c(x)$ is a slowly-varying function. In terms of operators, the dispersion operator is given by
\begin{subequations}
	\begin{align}
		\oper{D}_0 
			= 	&	\, [\hat{p}^\mu -q A^\mu_{\rm bg}(\hat{x})	]
					[ \hat{p}_\mu -q A_{\mathrm{bg},\mu}(\hat{x})] -m^2,		\\
	\oper{D}_{\rm osc} 
			= 	&	-  \{ q A^\mu_{\rm osc}(\hat{x})[ \hat{p}_\mu -q A_{\mathrm{bg},\mu}(\hat{x})] 	
					+ \hc	 \} \notag	\\
				&	
					+ q^2 A^\mu_{\rm osc}(\hat{x}) A_{\mathrm{osc},\mu}(\hat{x}).
	\end{align}
\end{subequations}
The corresponding Weyl symbols are (\App{app:Weyl})
\begin{subequations}	\label{eq:KG_D}
	\begin{align}
		D_0(x,p) 	=	&	\, \pi^2 -m^2,  
						\label{eq:KG_D0}		 	\\
		D_1(x,p)	= 	&	-2 q \pi \cdot A_{\rm osc}(x) ,
						\label{eq:KG_D1}			\\
		D_2(x,p)	= 	&	\, q^2 A_{\rm osc}(x) \cdot A_{\mathrm{osc}}(x),
						\label{eq:KG_D2}
	\end{align}
\end{subequations}
where $\pi^\mu(x,p) \doteq p^\mu -q A^\mu_{\rm bg}(x)$. Substituting \Eqs{eq:KG_D} into \Eq{eq:Deff_final}, we obtain
\begin{align}
	D_{\rm eff}(x,p)	
			=		&		\, \pi^2 - m^2 + \frac{q^2 |A_c|^2 }{2} \notag \\
					&		 - 	\sum_{n = \pm 1}
									\frac{q^2 | A_c \cdot (\pi+n  K/2) |^2}
										{2n	\pi 	\cdot K + n^2 K\cdot K },
\end{align}
where $|A_c|^2 = A_c \cdot A_c^*=|V_c|^2-|\vec{A}_c|^2$.

Following the procedure in \Sec{sec:pondero}, we determine the effective Hamiltonian for a point particle. Introducing the missing $c$ and $\hbar$ factors, we obtain
\begin{multline}
	H_{\rm eff}(t,\vec{X},\vec{P})	=  \,
				\gamma m c^2  +q V_{\rm bg} 
				-\frac{q^2 |A_c|^2 }{4 \gamma m c^2  } \\
				+\frac{1}{2\gamma m c^2 }
					\sum_{n = \pm 1}
						\frac{		q^2		| A_c  \cdot (\Pi_*+n\hbar K/2)   |^2}
						{2n \Pi_* \cdot \hbar K + n^2 \hbar^2 K \cdot K },
	\label{eq:KG_Heff}
\end{multline}
where
\begin{equation}
	\gamma(t,\vec{X},\vec{P}) \doteq 
				\sqrt{ 1+ 	\left(	\frac{\vec{P}}{mc}-	\frac{q\vec{A}_{\rm bg}}{mc^2} \right)^2   },
	\label{eq:KG_gamma}
\end{equation}
is the unperturbed Lorentz factor, $\smash{\Pi^\mu_* \doteq \left(	\gamma m c , \vec{P}-q\vec{A}_{\rm bg} /c	\right)}$ is the unperturbed kinetic four-momentum, $A^\mu_c(t, \vec{X}) = (V_c,\vec{A}_c) $ is the modulated four-potential, and $K^\mu(t,\vec{X})	= (\Omega/c, \vec{K})$ is the MW four-wavevector. All quantities are evaluated at the particle position $\vec{X}(t)$. 

Several interesting limits can be studied with the effective Hamiltonian \eq{eq:KG_Heff}. In the Lorentz gauge, where $\pd_\mu A^\mu_{\rm osc}=0$, we have $K \cdot  A_c = \mcu{O}(\epsilon)$. Then, $H_{\rm eff}$ becomes
\begin{multline}
	H_{\rm eff}(t,\vec{X},\vec{P})	=		
				\gamma m c^2  +q V_{\rm bg} 
				-\frac{q^2 |A_c|^2 }{4 \gamma mc^2  }  \\
				-\left( \frac{q^2		| A_c  \cdot \Pi_*   |^2}{\gamma mc^2 }	\right)
				\frac{K\cdot K}
				{4 (\Pi_* \cdot  K)^2 -  (\hbar K \cdot K)^2 }.
	\label{eq:KG_Heff2}
\end{multline}
Let us analyze the terms appearing in \Eq{eq:KG_Heff2}.  For example, $K \cdot K = (\Omega/c)^2 - \vec{K} = 0$ for a vacuum wave, so the second line vanishes. The remaining terms can be understood as the lowest-order expansion (in $|A_c|^2$) of the effective ponderomotive Hamiltonian $H_{\rm eff} = mc^2 [1 + \vec{\Pi}^2/(mc)^2 - q^2|A_c|^2/(2m^2c^4) ]^{1/2} + q V_{\rm bg}$ that a relativistic spinless particle experiences in an oscillating EM pulse \cite{Akhiezer:1956wb,Kibble:1966fj,Dodin:2003hz,Ruiz:2015dh}. In the case where $K \cdot K \neq 0$, the term in the second line of \Eq{eq:KG_Heff2} persists and accounts for Compton scattering, much like the Bragg scattering discussed in \Sec{sec:schro}.

Also, let us consider a particle that interacts with an oscillating electrostatic field so that $A_c^\mu=(V_c,0)$. In this case, \Eq{eq:KG_Heff} gives (\App{app:auxiliary})
\begin{multline}
	H_{\rm eff}(t,\vec{X},\vec{P})	=
		\gamma m c^2 + q V_{\rm bg} 
		+\frac{q^2 |V_c|^2}{4\gamma m }	 \\
			\times  \frac{	\vec{K}^2 -  (\vec{v}_* \cdot \vec{K}/c)^2 
		   				 - ( \hbar \vec{K} / 2 \gamma mc )^2 ( K \cdot K ) }
				{ ( \Omega - \vec{v}_* \cdot \vec{K})^2 -  (\hbar K \cdot K/ 2 \gamma m)^2 }.
	\label{eq:KG_GM}
\end{multline}
where $\vec{v}_* \doteq \vec{\Pi}/ (\gamma m)$ is the unperturbed particle velocity. The last term in \Eq{eq:KG_GM} is the relativistic ponderomotive energy. [In the nonrelativistic limit, when $\gamma \simeq 1$ and $\hbar | \vec{K}| \ll m c$, \Eq{eq:KG_GM} reduces to \Eq{eq:schro_Heff}, as expected.] When quantum corrections are negligible, we obtain
\begin{multline}
	H_{\rm eff}(t,\vec{X},\vec{P})	=
		\gamma m c^2 + q V_{\rm bg} 
		+\frac{q^2 |\vec{K}V_c|^2}{4 M ( \Omega - \vec{v}_* \cdot \vec{K})^2 }	,
\end{multline}
where $M \doteq m \gamma |\vec{K}|^2 \,  / \, [ \, |\vec{K}|^2 -  (\vec{v}_* \cdot \vec{K}/c)^2 ]$. When $\vec{v}_*$ is pointed along $\vec{K}$, one has $M = m\gamma^3$, which is understood as the longitudinal mass. In contrast, when $\vec{v}$ is transverse to $\vec{K}$, one has $M = m\gamma$, which is understood as the transverse mass \cite{Okun:1989jg}.

%%%%%%%%%%%%%%%%%%%%%%%%%%%%%%%%%%%%%%%%
\subsection{Example 3: Electrostatic wave in a density modulated plasma}
\label{sec:EM_wave}

As another example, let us consider an EM wave $\Psi(x)$ propagating in a density-modulated plasma. The PW dynamics is described by
\begin{equation}
	\pd_t^2 \Psi 	=  \del^2 \Psi -\omega_p^2 \Psi,
	\label{eq:EM_equation}
\end{equation}
where $\omega^2_p(x)	\doteq  4\pi q^2 n(x)/m $ is the plasma frequency squared \cite{Stix:1992waves}. The plasma density is modulated such that $n(x) = n_{\rm bg}(x) +  n_{\rm osc}(x) $, where $n_{\rm bg}(x)$ is the slowly-varying background plasma density and $\smash{ n_{\rm osc}(x) \doteq \mathrm{Re} [ n_c(x) e^{i\Theta(x)}  ]}$ is a fast modulation of small amplitude. The dispersion operator is given by
\begin{equation}
	\oper{D} = \hat{p} \cdot \hat{p} - \omega^2_{p}(\hat{x}),
\end{equation}
so the corresponding Weyl symbols are
\begin{subequations}	\label{eq:EM_D}
	\begin{align}
			D_0(x,p) 			=  &	\, p \cdot p  - \omega^2_{p,\mathrm{bg}}(x), \\
			D_{\rm osc}(x)  =	&	-\mathrm{Re} [  
										\omega^2_{p,c}(x) e^{i\Theta(x) }	],
		\end{align}
\end{subequations}
where $\omega^2_{p,\mathrm{bg}}(x) \doteq 4\pi q^2 n_{\rm bg} /m$ and $\omega^2_{p,c}(x) \doteq 4\pi q^2 n_c /m$.

Substituting \Eqs{eq:EM_D} into \Eq{eq:Deff_final}, we obtain
\begin{equation}
	D_{\rm eff}(x,p) =   
					p \cdot p - \omega_{p,\mathrm{bg}}^2  
					+ \frac{   |\omega^2_{p, c } |^2 
					 (K \cdot K) /8  }
								{ ( p \cdot K )^2
											- (K \cdot K)^2/4 } .
	\label{eq:EM_Deff}
\end{equation}
Then, the effective Hamiltonian is given by
\begin{equation}
	H_{\rm eff}(t,\vec{x},\vec{k}) = 
					\omega_0 (t,\vec{x},\vec{k}) 
					- \frac{|\omega^2_{p, \mathrm{c}}|^2}{16 \omega_0^3}\, \Xi ,
	\label{eq:EM_Heff}
\end{equation}
where $\omega_0 (t,\vec{x},\vec{k}) \doteq ( c^2\vec{k}^2+ \omega_{p,\mathrm{bg}}^2)^{1/2}$ is the unperturbed EM wave frequency, $\Xi$ is a dimensionless factor given by
\begin{equation}
	\Xi	\doteq \frac{ \Omega^2 - c^2 \vec{K}^2 }
		{  (  \Omega-\vec{v}_* \cdot \vec{K} )^2  - (\Omega^2-c^2\vec{K}^2)^2/4 \omega_0^2 },
	\label{eq:weird_fraction}
\end{equation}
and $\vec{v}_* = c^2 \vec{k}/ \omega_0$ is the unperturbed EM wave group velocity. (We have reintroduced the missing $c$ factors for clarity.) The second term in \Eq{eq:EM_Heff} represents the ponderomotive frequency shift that a classical EM wave experiences in a modulated plasma.

Similarly to the previous examples, the denominator in \Eq{eq:EM_Heff} also contains photon recoil effects. The Bragg resonance condition is also included; \ie for EM waves propagating in static modulated media $(\Omega=0)$, the Bragg resonance occurs at $2\vec{k}\cdot \vec{K} = \pm \vec{K}^2$. In the opposite limit where $\Omega \gg \vec{v}_* \cdot \vec{K}$, \Eq{eq:EM_Heff} becomes
\begin{equation}
	H_{\rm eff}(t,\vec{x},\vec{k}) = 
					\omega_0
					+ \left( \frac{|\omega^2_{p, \mathrm{c}}|^2}{16 \omega_0^3} \right)
					\frac{ N^2 - 1 }
								{  1 - \mu^2 (N^2-1)^2 },
	\label{eq:EM_Heff2}
\end{equation}
where $ \mu \doteq \Omega / 2\omega_0$ and $N \doteq  c |\vec{K}| /\Omega$ is the MW refraction index. Note that the GO result reported in \Ref{Dodin:2014iz} is recovered in the limit $ \mu  \ll 1$.

%%%%%%%%%%%%%%%%%%%%%%%%%%%%%%%%%%%%%%%
\section{Modulational dynamics and polarizability of wave quanta}
\label{sec:mod}

%--------------------------------------
\subsection{Basic equations}

Knowing the effective Hamiltonian $H_{\rm eff}$ of PWs, one can derive, without even considering the ray equations, the self-consistent dynamics of a MW when it interacts with an ensemble of PWs. (As a special case, when PWs are free charged particles, such ensemble is a plasma.) To do this, let us consider the action of the whole system in the form $\Lambda_\Sigma = \Lambda_{\rm mw} + \Lambda_{\rm pw}$, where $\Lambda_{\rm mw}$ is the MW action and $\Lambda_{\rm pw}$ is the cumulative action of all PWs. We attribute the interaction action to $\Lambda_{\rm pw}$, so, by definition, $\Lambda_{\rm mw}$ is the system action absent PWs. Then, $\Lambda_{\rm mw}$ equals the action of the MW EM field in vacuum, $\smash{\Lambda_{\rm mw} \approx \int \mathrm{d}^4x \,(\vec{E}_{\rm mw}^2 - \vec{B}_{\rm mw}^2)/(8\pi)}$ \cite{Goldstein:2002up}. Since the MW is assumed to satisfy the GO approximation, its electric and magnetic fields can be expressed as
\begin{equation}
	\vec{E}_{\rm mw} = \text{Re}\,(\vec{E}_c e^{i\Theta}), 
	\quad
	\vec{B}_{\rm mw} = \text{Re}\,(\vec{B}_c e^{i\Theta}), 
\end{equation}
where the envelopes $\vec{E}_c$ and $\vec{B}_c$ are slow compared to $\Theta$. Then, we can approximate $\Lambda_{\rm mw}$ as
\begin{equation}
	\Lambda_{\rm mw} = \int \mathrm{d}^4x\,
									\left(\frac{|\vec{E}_c|^2}{16\pi} - \frac{c^2|\vec{K} \times \vec{E}_c|^2}{16\pi \Omega^2}\right),
\end{equation}
where we substituted Faraday's law $\vec{B}_c \approx (c\vec{K}/\Omega) \times \vec{E}_c$.

To calculate $\Lambda_{\rm pw}$, we assume that PWs are mutually incoherent and do not interact other than via the MW. (When PWs are charged particles, this is known as the collisionless-plasma approximation. In a broader context, this can be recognized as the quasilinear approximation; for instance, see \Ref{Ruiz:2016gv}.) Then, $\Lambda_{\rm pw} = \sum_i \Lambda_i$, where $\Lambda_i$ are the actions of the individual PWs. Adopting $\Lambda_i$ in the form of \Eq{eq:action_Hayes}, we obtain
\begin{equation}\label{eq:LambdaP}
	\Lambda_{\rm pw} = \Lambda_{\rm pw,0} 
									- \sum_i \int \mathrm{d}^4x\, \mc{I}_i(t, \vec{x})\,\Phi_i(t, \vec{x}, \del \theta_i),
\end{equation}
where $\smash{ \Lambda_{\rm pw,0}= - \sum_i \int \mathrm{d}^4x\, \mc{I}_i \, [ \pd_t \theta_i + H_{0,i}(t,\vec{x},\del \theta_i)]}$ is independent of the MW variables, so it can be dropped. (In this section, we are only interested in ELEs for the MW, and $\Lambda_{\rm pw,0}$ does not contribute to those.) Let us consider PWs in groups $s$ such that, within each group, PWs have the same ponderomotive frequency shift $\Phi_s$. Then, we can rewrite $\Lambda_{\rm pw}$ as follows:
\begin{equation}\label{eq:LambdaP2}
	\Lambda_{\rm pw} = - \sum_s \int \mathrm{d}^4x \,\mathrm{d}^3 \vec{p} \,
										f_s(t, \vec{x}, \vec{p})\,\Phi_s(t, \vec{x}, \vec{p}),
\end{equation}
where $f_s \doteq \sum_{i \in s} \mc{I}_i(t, \vec{x})\,\delta[\vec{p} - \del\theta_i(t, \vec{x})]$. This gives $\Lambda_\Sigma = \int \mathrm{d}^4x\,\mcc{L}$, where the Lagrangian density $\mcc{L}$ is
\begin{equation}\notag
	\mcc{L} = \frac{|\vec{E}_c|^2}{16\pi} 
					- \frac{c^2|\vec{K} \times \vec{E}_c|^2}{16\pi \Omega^2} 
					-  \sum_s \! \int \! \mathrm{d}^3 \vec{p} \, f_s(t, \vec{x}, \vec{p})\,\Phi_s(t, \vec{x}, \vec{p}).
\end{equation}

The meaning of $f_s$ is understood as follows. A single wave with a well-defined local momentum $\del\theta_i(t, \vec{x})$ has a phase-space distribution that is delta-shaped along the local wavevector (momentum) coordinate, $\propto \delta[\vec{p} - \del\theta_i(t, \vec{x})]$. The coefficient in front of the delta function must be the spatial probability density for the proper normalization. In our case, the spatial probability density is $|\psi_i(t, \vec{x})|^2$, which is the same as $\mc{I}_i(t, \vec{x})$. Thus, $\mc{I}_i(t, \vec{x})\,\delta[\vec{p} - \del\theta_i(t, \vec{x})]$ is the phase-space density of $i$th wave. This makes $f_s(t,\vec{x},\vec{p})$ the total phase-space density of species $s$. This formulation can also be applied, for example, to degenerate plasmas to the extent that the Hartree approximation is applicable \cite{Lifshitz:1981ui,book:Haas}. Specifically, if the spin-orbital interaction is negligible and particles interact with each other only through the mean EM field, their Lagrangian densities sum up (by definition of the mean-field approximation), so one recovers the same $\mcc{L}$ as in nondegenerate plasma. The only subtlety in this case is that $f_s(t,\vec{x},\vec{p})$ is now restricted by Pauli's exclusion principle (or, in equilibrium, to Fermi-Dirac statistics).

Since $\Phi_s$ is bilinear in the MW field and independent on the MW phase, it can be expressed as
\begin{equation}\label{eq:Phialpha}
	\Phi_s = - \frac{1}{4} \,\vec{E}_c^*\cdot \vec{\alpha}_s \cdot \vec{E}_c,
\end{equation}
where $\vec{\alpha}_s$ is some complex tensor that can depend on $\Omega$ and $\vec{K}$ but not on $\vec{E}_c$ or $\vec{E}_c^*$. Explicitly, it is defined~as
\begin{equation}\label{eq:alpha}
	\vec{\alpha}_s \doteq -4\, \frac{\pd^2}{\pd\vec{E}_c\,\pd\vec{E}_c^*}\,
										\Phi_s (\vec{E}_c, \vec{E}_c^*, \Omega, \vec{K}; \,t, \vec{x}, \vec{p}),
\end{equation}
or, equivalently, $\vec{\alpha}_s \doteq -4\pd^2 H_{{\rm eff}, s}/(\pd\vec{E}_c\,\pd\vec{E}_c^*)$. Then, 
\begin{equation}
	\mcc{L} = \frac{1}{16\pi}\,\vec{E}_c^* \cdot \vec{\varepsilon}(t, \vec{x}, \Omega, \vec{K}) \cdot \vec{E}_c 
					- \frac{c^2|\vec{K} \times \vec{E}_c|^2}{16\pi \Omega^2},
\end{equation}
where we introduced $\vec{\varepsilon} \doteq 1 + \vec{\chi}$ and 
\begin{equation}
	\vec{\chi} \doteq 4\pi \sum_s \int \mathrm{d}^3 \vec{p} \,
												f_s(t, \vec{x}, \vec{p})\,
												\vec{\alpha}_s(t, \vec{x}, \vec{p}, \Omega, \vec{K}).
	\label{eq:chi}
\end{equation}

By treating $(\Theta, \vec{E}_c, \vec{E}_c^*)$ as independent variables, we then obtain the following ELEs:
\begin{align}
	\delta \Theta: 			&	\quad \pd_t (\pd_\Omega \mcc{L}) - \del \cdot (\pd_\vec{K} \mcc{L}) = 0,\\
	\delta \vec{E}_c^*:	&	\quad (\Omega/c)^2\,\vec{\varepsilon} \cdot \vec{E}_c 
													+ \vec{K} \times (\vec{K} \times \vec{E}_c) = 0,
\end{align}
plus a conjugate equation for $\vec{E}_c^*$. [Remember that $\Omega$ and $\vec{K}$ are related to $\Theta$ via \Eq{eq:MW_Omega_K}.] We then recognize these ELEs as the GO equations describing EM waves in a dispersive medium with dielectric tensor $\vec{\varepsilon}$ \cite{Dodin:2012hn, Stix:1992waves}. Thus, $\vec{\chi}$ is the susceptibility of the medium, and $\vec{\alpha}_s$ serves as the \textit{linear polarizability} of PWs of type $s$ \cite{foot:non_GO}. 

Hence, \Eq{eq:Phialpha} can be interpreted as a fundamental relation between the ponderomotive energy and the linear polarizability. This relation represents a generalization of the well-known ``$K$-$\chi$ theorem'' \cite{Cary:1977ud,Kaufman:1987aa,Dodin:2010jk}, which establishes this equality for classical particles, to general waves. Those include quantum particles as a special case and also photons, plasmons, phonons, etc. According to the theory presented here, any such object can be assigned a ponderomotive energy and thus has a polarizability \eq{eq:alpha}. Some examples are discussed below.

%--------------------------------------
\subsection{Examples}

As a first example, let us consider a nonrelativistic quantum electron with charge $q$, mass $m$, and OC momentum $\vec{P} \doteq m\vec{v}$. Suppose the electron interacts with an electrostatic MW (so $\vec{B}_c = 0$). Then, $H_{\rm eff}$ can be taken from \Eq{eq:schro_Heff}, and \Eq{eq:alpha} readily yields that the electron polarizability is a diagonal matrix given by
\begin{equation}\label{eq:alphae}
	\vec{\alpha}_e = - \mathbb{I}_3\,\frac{q^2}{m} 
							\left[(\Omega - \vec{K} \cdot \vec{v})^2 - (\hbar \vec{K}^2 /2m)^2\right]^{-1}.
\end{equation}
(Here, $\mathbb{I}_3$ is a $3 \times 3$ unit matrix.) From \Eq{eq:chi}, the susceptibility of the electron plasma is given by \cite{foot:resonance}
\begin{equation}
	\vec{\chi} = - \mathbb{I}_3\,\frac{4 \pi q^2}{m}  \int \mathrm{d}^3 \vec{v}\, 
						\frac{f(t, \vec{x}, \vec{v})}{(\Omega - \vec{K} \cdot \vec{v})^2 - (\hbar \vec{K}^2 /2 m)^2},
\end{equation}
which is precisely the textbook result \cite{Lifshitz:1981ui,book:Haas}. This shows that the commonly known expression for the dielectric tensor of quantum plasmas is actually a reflection of the less-known quantum ponderomotive energy.

Second, consider an EM wave in a nonmagnetized density-modulated cold electron plasma. Using Gauss's law, one readily finds that $	\omega_{p,c}^2 = (iq/m)\vec{K} \cdot \vec{E}_c$, where we assume the same notation as in \Sec{sec:EM_wave}. Then, using \Eq{eq:EM_Heff}, one gets
\begin{equation}
	\Phi = - \frac{\hbar q^2 \Xi}{16 m^2\omega_0^3}\,(\vec{E}_c^*\cdot \vec{K}  \vec{K} \cdot \vec{E}_c),
\end{equation}
where $\vec{K} \vec{K}$ is a dyadic tensor, and $\Xi$ is given by \Eq{eq:weird_fraction}. (The factor $\hbar$ is introduced in order to treat $\Phi$ as a per-photon energy rather than as a classical frequency.) Hence, \Eq{eq:alpha} gives that the photon polarizability is
\begin{equation}\label{eq:aph}
	\vec{\alpha}_{\rm ph} = \frac{\hbar q^2 \Xi}{4m^2\omega_0^3}\,\vec{K}  \vec{K}.
\end{equation}
In principle, one must account for this polarizability when calculating $\vec{\varepsilon}$; \ie photons contribute to the linear dielectric tensor just like electrons and ions \cite{foot:photon,Bingham:1997aa}. That said, the effect is relatively small, and ignoring the photon contribution to the plasma dielectric tensor is justified except at large enough photon densities \cite{foot:photon}.

Similar calculations are also possible for dissipative dynamics and vector waves and also help understand the modulational dynamics of wave ensembles in a general context. However, elaborating on these topics is outside the scope of this paper and is left to future publications.

%%%%%%%%%%%%%%%%%%%%%%%%%%%%%%%%%%%%%%%%%%%%%%
%%%%%%%%%%%%%%%%%%%%%%%%%%%%%%%%%%%%%%%%%%%%%%
\section{Conclusions}
\label{sec:conclusions}

In this work, we show that scalar waves, both classical and quantum, can experience time-averaged refraction when propagating in modulated media. This phenomenon is analogous to the ponderomotive effect encountered by charged particles in high-frequency EM fields. We propose a covariant variational theory of this ``ponderomotive effect on waves'' for a general nondissipative linear medium. Using the Weyl calculus, our formulation is able to describe waves with temporal and spatial period comparable to that of the modulation (provided that parametric resonances are avoided). This theory can be understood as a generalization of the oscillation-center theory, which is known from classical plasma physics, to any linear waves or quantum particles in particular. This work also shows that any wave is, in fact, a polarizable object that contributes to the linear dielectric tensor of the ambient medium. Three examples of applications of the theory are given: a Schr\"odinger particle propagating in an oscillating electrostatic field, a Klein--Gordon particle interacting with modulated EM fields, and an EM wave propagating in a density-modulated plasma. 

This work can be expanded in several directions. First, one can extend the theory to dissipative waves \cite{foot:dissipation} and vector waves with polarization effects \cite{Ruiz:2015hq,Littlejohn:1991vj}, which could be important at Bragg resonances. Second, the theory presented here can be used as a stepping stone to improving the understanding of the modulational instabilities in general wave ensembles. This requires a generalization of the analysis presented in \Sec{sec:mod} and will be reported in a separate paper. 

The authors thank N.~J.~Fisch for valuable discussions. This work was supported by the U.S. DOE through Contract No. DE-AC02-09CH11466, by the NNSA SSAA Program through DOE Research Grant No. DE-NA0002948, and by the U.S. DOD NDSEG Fellowship through Contract No. 32-CFR-168a.

\appendix

%%%%%%%%%%%%%%%%%%%%%%%%%%%%%%%%%%%%%%%%%%%%%%
%%%%%%%%%%%%%%%%%%%%%%%%%%%%%%%%%%%%%%%%%%%%%%
\section{The Weyl transform}
\label{app:Weyl}

This appendix summarizes our conventions for the Weyl transform. (For more information, see the excellent reviews in \Refs{Imre:1967fr,McDonald:1988dp,BakerJr:1958bo,Tracy:2014to}.) The Weyl symbol $A(x,p)$ of any given operator $\oper{A}$ is defined as
\begin{equation}
	A(x,p) \doteq 
				\int \mathrm{d}^4 s \, e^{i p \cdot s }
				\braket{ x+ s/2 | \oper{A}	| x	-	s/2	}  ,
	\label{def:weyl_symbol}
\end{equation}
where $p \cdot s = p_0 s_0 - \vec{p}\cdot \vec{s}$ and the integrals span over $\mathbb{R}^4$. We shall refer to this description of the operators as a \textit{phase-space representation} since the symbols \eq{def:weyl_symbol} are functions of the eight-dimensional phase space. Conversely, the inverse Weyl transformation is given by
\begin{equation}
	\oper{A} = \int	
					\frac{\mathrm{d}^4 x \,  \mathrm{d}^4 p \,  	\mathrm{d}^4 s }{(2\pi)^4} \,
					e^{i p \cdot s } 
					A(x,p) \ket{x-s/2} \bra{x+s/2}.
	\label{eq:weyl_inverse}
\end{equation}
Hence, $\mcu{A}(x,x') = \braket{x|\oper{A}|x'}$ can be expressed as
\begin{equation}
	\mcu{A}(x,x') =  \int
 					\frac{\mathrm{d}^4 p}{(2\pi )^4} \,
					e^{i p \cdot (x'-x)  }
					A		\left(	\frac{x+x'}{2},	p	\right).
	\label{eq:weyl_x_rep}
\end{equation}

In the following, we will outline a number of useful properties of the Weyl transform.

\begin{itemize}[leftmargin=*]
\item For any operator $\oper{A}$, the trace $\mathrm{Tr}[\oper{A}] \doteq \int \mathrm{d}^4 x \, \braket{ x | \oper{A} | x }$ can be expressed as
\begin{equation}
	\mathrm{Tr}[\oper{A}] 
				=  \int \frac{\mathrm{d}^4 x \,  \mathrm{d}^4 p }{(2\pi  )^4} \,  
				A(x, p ).
	\label{eq:trace}
\end{equation}

\item If $A(x,p)$ is the Weyl symbol of $\oper{A}$, then $A^*(x,p)$ is the Weyl symbol of $\oper{A}^\dag$. As a corollary, the Weyl symbol of a Hermitian operator is real.

\item For any $\oper{C} =\oper{A} \oper{B}$, the corresponding Weyl symbols satisfy \cite{Moyal:1949gj,Groenewold:1946kp}
\begin{equation}
	C (x,p) = A (x,p) \star B (x,p).
	\label{eq:Moyal}
\end{equation}
Here `$\star$' refers to the \textit{Moyal product}, which is given by
\begin{equation}
		A(x,p) \star B (x,p) 
		\doteq 
		A (x,p) e^{i \hat{\mc{L}} /2 }  B (x,p),
	\label{def:Moyal}
\end{equation}
and $\hat{\mc{L}}$ is the \textit{Janus operator}, which is given by
\begin{equation}
	\hat{\mc{L}}  \doteq  
			\overleftarrow{\pd_p} \cdot \overrightarrow{\pd_x} -		
			\overleftarrow{\pd_x} \cdot \overrightarrow{\pd_p}
			=
			\{ {\cdot} , {\cdot} \}.
\end{equation}
The arrows indicate the direction in which the derivatives act, and $A \hat{\mc{L}} B = \{ A, B \}$ is the canonical Poisson bracket in the eight-dimensional phase space, namely,
\begin{equation}
	\hat{\mc{L}}  =
			\frac{\overleftarrow{\pd} }{\pd p^0}  \frac{ \overrightarrow{\pd} }{\pd x^0}
	-		\frac{\overleftarrow{\pd} }{\pd x^0}  \frac{ \overrightarrow{\pd} }{\pd p^0}
	+		\frac{\overleftarrow{\pd} }{\pd \vec{x}}  \cdot \frac{ \overrightarrow{\pd} }{\pd \vec{p} }
	-		\frac{\overleftarrow{\pd} }{\pd \vec{p}}  \cdot \frac{ \overrightarrow{\pd} }{\pd \vec{x} }.
\end{equation}

\item The Moyal product is associative; \ie 
\begin{equation}
	A \star B \star C 
			= (A \star B) \star C 
			= A \star (B \star C).
\end{equation}

\item The anti-symmetrized Moyal product defines the so-called \textit{Moyal bracket}
\begin{equation}
	\moysin{ A ,	B } 
	 		\doteq \frac{1}{i} \left(  A \star B - B \star A	\right) 
	 		= 2 A \sin \left( \frac{\hat{\mc{L}} }{2} \right) B.
	\label{eq:sine_bracket}
\end{equation}
Because of the latter equality, the Moyal bracket is also often called the \textit{sine bracket}. To the lowest order in $\epsilon$,
\begin{equation}
	\moysin{A,B} \simeq  \{A, B \} .
\end{equation}

\item Now we tabulate some Weyl transforms of various operators. (We use a two-sided arrow to show the correspondence with the Weyl transform.) First of all, the Weyl transforms of the identity, position, and momentum operators are given by
\begin{equation}
	\hat{1}	\,	\Leftrightarrow	\,	1, \quad
	\hat{x}^\mu \,	\Leftrightarrow	\, x^\mu, \quad
	\hat{p}_\mu \,	\Leftrightarrow	\, p_\mu.
\end{equation}
For any two functions $f$ and $g$, one has
\begin{equation}
	f(\hat{x}) 	\,	\Leftrightarrow	\, f(x), \quad
	g(\hat{p}) 	\,	\Leftrightarrow	\, g(p).
\end{equation}
Similarly, using \Eq{def:Moyal}, one has
\begin{gather}
	\hat{p}_\mu	f(\hat{x}) 	\,	\Leftrightarrow	\, p_\mu f(x)	+ (i/2) \pd_\mu 	f(x), \\
	f(\hat{x}) \hat{p}_\mu	\,	\Leftrightarrow	\, p_\mu f(x)	- (i/2) \pd_\mu 	f(x).
\end{gather}

\end{itemize}

%%%%%%%%%%%%%%%%%%%%%%%%%%%%%%%%%%%%%%%%%%%%%%
%%%%%%%%%%%%%%%%%%%%%%%%%%%%%%%%%%%%%%%%%%%%%%
\section{Auxiliary calculations}
\label{app:auxiliary}

Here we include additional details on some of the calculations that are reported in the main text. In particular, to obtain \Eq{eq:aux1}, we note that $\Theta(x)$ is fast while $K_\mu(x) \doteq -\pd_\mu \Theta$ is slow; then,
\begin{align}
A(x&,p) \star e^{i \Theta  }	
			\notag \\
	&	=	A(x,p) e^{i  \oper{L}/2}  e^{i \Theta  }
			\notag \\
	&	=	A(x,p) 
			\left( \sum_{n=0}^\infty \frac{i^n  }{2^n n!}  
				\frac{\overleftarrow{\pd}^n}{\pd p^n} \cdot \frac{\overrightarrow{\pd}^n}{\pd x^n}
			\right)	
			e^{i \Theta }
			\notag \\	
	&	=	A(x,p) 
			\left[ \sum_{n=0}^\infty \frac{ i^n   }{2^n n!}  
				\frac{\overleftarrow{\pd}^n}{\pd p^n}  \cdot
				\left( i\frac{\pd \Theta}{\pd x}	\right)^n
			\right]	
			e^{i \Theta  }
			+\mcu{O}(\epsilon_{\rm mw})
			\notag \\		
	&	=	A(x,p) 
			\left[ \sum_{n=0}^\infty \frac{ 1 }{n!}  
				\frac{\overleftarrow{\pd}^n}{\pd p^n}	\cdot
				\left( \frac{K}{2}	\right)^n
			\right]	
			e^{i \Theta  }
			+\mcu{O}(\epsilon_{\rm mw})
			\notag \\
	&	=	A(x,p+K/2) 
			e^{i \Theta }
			+\mcu{O}(\epsilon_{\rm mw}),
\end{align}
where the symbol `$\cdot$' denotes contraction. Similarly,
\begin{equation}
	e^{i \Theta  } \star	A(x,p)  
			= 			A(x,p-K/2) e^{i \Theta  }
			+\mcu{O}(\epsilon_{\rm mw}).
\end{equation}

For the calculation shown in \Eq{eq:C2_aux}, we need the following result:
\begin{align}
	A&(x,p) 	 e^{i \Theta_1 } \star  B(x,p) e^{  i \Theta_2  }
							\notag \\
				&	=		A(x,p) 	 
							e^{i \Theta_1} 
							e^{i  \oper{L}/2}  
							B(x,p) 
							e^{i \Theta_2  }
							\notag \\
				&	=		A(x,p) 	
							e^{i \Theta_1  } e^{i 
									( \overleftarrow{\pd_p} \cdot \overrightarrow{\pd_x} -		
											\overleftarrow{\pd_x} \cdot \overrightarrow{\pd_p})/2}  
							e^{i \Theta_2 } 							
							B(x,p) 
							\notag \\
				&	=		A(x,p) 	
							  e^{i \Theta_1  } e^{-
									( \overleftarrow{\pd_p} \cdot \pd_x \Theta_2 -		
											\pd_x \Theta_1 \cdot \overrightarrow{\pd_p})/2}  
							e^{i \Theta_2 } 						
							B(x,p)
							\notag \\ 
				&			~~~+\mcu{O}(\epsilon_{\rm mw})
							\notag \\
				&	=		A(x,p) 	
							e^{i \Theta_1  } 
							e^{\overleftarrow{\pd_p} \cdot (K_2 /2)}
							e^{-(K_1/2) \cdot \overrightarrow{\pd_p}}  			
							e^{i \Theta_2  } 				
							B(x,p) 
							\notag \\ 
				&			~~~+\mcu{O}(\epsilon_{\rm mw})
							\notag \\
				&	=		A(x,p+K_2/2)
							B(x,p-K_1/2) 	
							e^{i (\Theta_1+ \Theta_2)  } 
							+\mcu{O}(\epsilon_{\rm mw}).
\end{align}
Substituting this result, we then obtain
\begin{align}
	C_2
		&	= 	\moysin{ T_1, D_1 }/2
				\notag \\
		&	=	\moysin{ 	\mc{T}_1(x,p)e^{i\Theta}, 
								\mc{D}_1^*(x,p) e^{-i\Theta}  }/8
				\notag \\
		&		~~~~	+
				\moysin{ 	\mc{T}_1^*(x,p) 	e^{-i\Theta}, 
								\mc{D}_1(x,p) 		e^{i\Theta}  }/8
				\notag \\ 
		&		~~~~	+
				\moysin{ 	\mc{T}_1(x,p)e^{i\Theta}, 
								\mc{D}_1(x,p) e^{i\Theta}  }/8
				\notag	\\
		&		~~~~ 	+
				\moysin{ 	\mc{T}_1^*(x,p) e^{-i\Theta}, 
								\mc{D}_1^*(x,p) e^{-i\Theta}  }/8
				\notag \\ 
		&	=	\mc{T}_1(x, p -K/2) \mc{D}_1^*(x, p - K/2) / (8i)					\notag\\
		&		~~~~ - \mc{T}_1(x, p + K /2) \mc{D}_1^*(x, p + K/2) / (8i)		\notag\\
		&		~~~~ + \mc{T}_1^*(x, p + K /2) \mc{D}_1(x, p + K/2) / (8i)		\notag\\
		&		~~~~ - \mc{T}_1^*(x, p - K /2) \mc{D}_1(x, p - K/2) / (8i)			\notag \\
		& 		~~~~	+	\mathrm{Re} [ \mc{C}_2(x,p) e^{2i\Theta(x)} ] +\mcu{O}(\epsilon_{\rm mw}),
		\label{app:aux_eq}
\end{align}
where $\mc{C}_2(x,p)$ is some function, whose explicit expression is not important for our purposes. Substituting \Eq{eq:G} into \Eq{app:aux_eq}, we obtain
\begin{align}
	C_2
		&	= -\frac{1}{4}
				\bigg[
					\frac{  |\mc{D}_1(x, p + K/2)|^2}{ D_0(x, p+K) - D_0(x, p)|^2}	\notag \\
		&	~~~~~~~~~~~~~	
					 + \frac{  |\mc{D}_1(x, p - K/2)|^2}{ D_0(x, p-K) - D_0(x, p)}
				\bigg]
				\notag 		\\
		&	~~~~	
				+ \mathrm{Re} [ \mc{C}_2(x,p) e^{2i\Theta(x)} 	]
				+\mcu{O}(\epsilon_{\rm mw})
				\notag\\
		&	= -\frac{1}{4}
				\sum_{n = \pm 1}
					\frac{  |\mc{D}_1(x, p + n K/2)|^2}{ D_0(x, p+ nK) - D_0(x, p)|^2}
				\notag \\
		&	~~~~					
				+ \mathrm{Re} [ \mc{C}_2(x,p) e^{2i\Theta(x)} ]
				+\mcu{O}(\epsilon_{\rm mw}).
\end{align}

The calculation of \Eq{eq:KG_GM} is presented below. Starting from \Eq{eq:KG_Heff} and letting $A_c^\mu =(V_c,0)$, we have
\begin{widetext}
\begin{align}
	H_{\rm eff}(t,\vec{X}, & \vec{P}) 
					- \gamma m c^2  - q V_{\rm bg} 
					\notag \\	
		= 	&		-\frac{q^2 (A_c \cdot A_c^*) }{4 \gamma m  c^2 } 
					+\frac{1}{2\gamma m c^2 }
						\sum_{n = \pm 1}
							\frac{		q^2		| A_c  \cdot (\Pi_*+n\hbar K/2)   |^2}
										{2n \Pi_* \cdot \hbar K + n^2 \hbar^2 K \cdot K },
					\notag \\
		=	& 		-\frac{q^2 |V_c|^2 }{4 \gamma m c^2 } 
					+\frac{q^2 |V_c|^2}{2\gamma m c^2 }
						\left[
							\frac{		(\gamma m  c+  \hbar \Omega/2 c)^2 }
										{2 \Pi_* \cdot \hbar K + \hbar^2 K \cdot K }
							-
							\frac{		(\gamma m c - \hbar \Omega/2 c)^2 }
										{2 \Pi_* \cdot \hbar K - \hbar^2 K \cdot K }
						\right]
					\notag \\
		=	& 		-\frac{q^2 |V_c|^2 }{4 \gamma m c^2 } 
					+\left( \frac{q^2 |V_c|^2}{2\gamma m c^2 } \right)
					\frac{		
						(\gamma m c +  \hbar \Omega/2 c)^2 (2 \Pi_* \cdot \hbar K - \hbar^2 K \cdot K) 
					   -(\gamma m  c-  \hbar \Omega/2 c)^2 (2 \Pi_* \cdot \hbar K + \hbar^2 K \cdot K) }
										{4 (\Pi_* \cdot \hbar K)^2 - (\hbar^2 K \cdot K)^2 }
					\notag \\
		=	& 	\,	-\frac{q^2 |V_c|^2 }{4 \gamma m c^2 } 
					+ \left( \frac{q^2 |V_c|^2}{8\gamma m c^2 }	\right)
					\frac{		
						4 \gamma m  \Omega   ( \Pi_* \cdot  K ) 
					   -  2 \gamma^2 m^2 c^2 ( K \cdot K)    
					   - \hbar^2 \Omega^2 ( K \cdot K)/2c^2 }
										{(\Pi_* \cdot  K)^2 - (\hbar K \cdot K/2)^2  }
					\notag \\
		=	& 		\, \left( \frac{q^2 |V_c|^2}{8\gamma m c^2 }  \right)
					\frac{		
						4 \gamma m  \Omega ( \Pi_* \cdot  K )  
					   - 2  \gamma^2 m^2 c^2  ( K \cdot K)  
					   - 2 (\Pi_* \cdot  K)^2 
					   - \hbar^2 \Omega^2 ( K \cdot K)/2c^2
					   +\hbar^2 (K \cdot K)^2/2 }
						{(\Pi_* \cdot  K)^2 - (\hbar K \cdot K/2)^2  }
					\notag \\
		=	& 		\,  \left(  \frac{q^2 |V_c|^2}{8\gamma m c^2 }  \right)
					\frac{		
						4 \gamma m  \Omega   ( \gamma m \Omega - \vec{\Pi} \cdot \vec{K}) 
					   - 2  \gamma^2 m^2  c^2 ( \Omega^2 / c^2 - \vec{K}^2)  
					   - 2 ( \gamma m \Omega - \vec{\Pi} \cdot \vec{K})^2
					   - \hbar^2 \vec{K}^2 ( K \cdot K)/2 }
						{(\Pi_* \cdot  K)^2 -  (\hbar K \cdot K/2)^2 }
					\notag \\
		=	& 		\,  \left(  \frac{q^2 |V_c|^2}{4\gamma m } \right)
					\frac{		
						\vec{K}^2 -  (\vec{\Pi} \cdot \vec{K}/ \gamma m c)^2 
						 - ( \hbar \vec{K} / 2 \gamma mc )^2 ( K \cdot K ) }
						{ ( \Omega - \vec{\Pi} \cdot \vec{K}/ \gamma m )^2 -  (\hbar K \cdot K/ 2 \gamma m)^2 }
					\notag \\
		=	& 		\,  \left(  \frac{q^2 |V_c|^2}{4\gamma m }	\right)
					\frac{		
						\vec{K}^2 -  (\vec{\Pi} \cdot \vec{K}/ \gamma m c)^2  }
						{ ( \Omega - \vec{\Pi} \cdot \vec{K}/ \gamma m )^2  }
					+ \mcu{O} \left(  \hbar^2 	\right).
\end{align}

\end{widetext}

%%%%%%%%%%%%%%%%%%%%%%%%%%%%

%{ \color{blue} $D = D_0 + D_{\rm osc}$ }
%
%
%{ \color{blue} $D_{\rm eff} $ }
%
%
%{ \color{red} $\Psi $ }
%
%
%{ \color{red} $\psi$ }
%%
%%
%%(a)
%%
%%(b)


\begin{thebibliography}{71}
\bibitem[{foo({\natexlab{c}})}]{foot:sigma}
\bibinfo{note}{We will incorporate the factor $\sigma$ in the
  definition of $\hat{\mc{D}}_n$ eventually, but for now, it is convenient to have it
  written explicitly.}
\bibitem[{foo({\natexlab{d}})}]{foot:many_waves}
\bibinfo{note}{One can also easily generalize further calculations to the case
  when multiple incoherent GO MW are present.}
\bibitem[{foo({\natexlab{e}})}]{foot:slowly}
\bibinfo{note}{By a slowly-varying function $f(x,p)$ in spacetime, we mean an
  infinitely differentiable function that may be written as $f(\epsilon x, p)$
  so that its $n$th spacetime derivatives are $\mc{O}(\epsilon)$.}
\bibitem[{foo({\natexlab{f}})}]{foot:quasi}
\bibinfo{note}{By a quasi-periodic function $f(x,p)$ in spacetime, we mean one
  that may be written as $g(\epsilon x , p, \Theta)$, where $g(a, p, \Theta)$
  is $2 \pi$-periodic for any constant $a$ and where the phase $\Theta$ is a
  function such that its derivatives are nonzero and slowly-varying.}
\bibitem[{foo({\natexlab{j}})}]{foot:non_GO}
\bibinfo{note}{Alternatively, one can redo the calculation in \Sec{sec:mod}
  with $\Lambda_{\rm pw}$ in the form \eq{eq:action_eff}. Then, $\vec{\chi}$ is
  conveniently expressed as $\vec{\chi} = 4\pi \sum_s \int \mathrm{d}^4p \,
  F_s(x, p) \, \vec{\mc{A}}_s(x, p, K)$, where $F_s \doteq \sum_{i \in s} W_i$
  the sum of the Wigner functions of species $s$, and $\vec{\mc{A}}$ is the
  generalized polarizability defined via $D_{\rm NL} = (1/4)\,\vec{E}_c^* \cdot
  \vec{\mc{A}} \cdot \vec{E}_c$, where $D_{\rm NL}$ is the MW-dependent part of
  $D_{\rm eff}$.}
\bibitem[{foo({\natexlab{k}})}]{foot:resonance}
\bibinfo{note}{Strictly speaking, one must only take the principal value of the integral of the susceptibility since we assumed that the dielectric tensor $\vec{\varepsilon}$ is Hermitian. For a rigorous treatment of the resonant denominator, which gives rise to Landau damping, see the discussion in \Ref{foot:dissipation}.}
\end{thebibliography}
\end{document}